\documentclass[%
aip,
amsmath,amssymb,
preprint,%
]{revtex4-1}

\usepackage{graphicx}
\usepackage{dcolumn}
\usepackage{bm}
\usepackage{subcaption}

\usepackage[utf8]{inputenc}
\usepackage[T1]{fontenc}
\usepackage{mathptmx}
\usepackage{etoolbox}
\usepackage{comment}

\makeatletter
\def\@email#1#2{%
	\endgroup
	\patchcmd{\titleblock@produce}
	{\frontmatter@RRAPformat}
	{\frontmatter@RRAPformat{\produce@RRAP{*#1\href{mailto:#2}{#2}}}\frontmatter@RRAPformat}
	{}{}
}%
\makeatother
\begin{document}
	
	
	\title[Evaluation of the Ambipolar Diffusion Approximation in Partially Ionized Rarefied Hypersonic Flows]{Evaluation of the Ambipolar Diffusion Approximation in Partially Ionized Rarefied Hypersonic Flows}
	\author{Marisa Petrusky}
	\email{marisa.petrusky@colorado.edu}
	\author{Iain D. Boyd}%
	\affiliation{ 
		Ann and H.J. Smead Department of Aerospace Engineering Sciences,
		University of Colorado, Boulder, Colorado, 80309, USA
	}%
	
	\date{\today}
	
	\begin{abstract}
		
		Accurate numerical simulation of rarefied hypersonic plasmas is increasingly important for optimization of re-entry spacecraft design and the development of advanced aerospace technologies. For kinetic simulation methods, it is convention to enforce ions and electrons to diffuse at the same rate, known as the ambipolar diffusion approximation. This approach circumvents costly resolution of fast electron motion, but neglects the complex plasma dynamics of ions and electrons. Almost all studies that investigated the efficacy of the ambipolar diffusion approximation in hypersonics report noticeable differences in flowfield properties when electrostatic modeling is used, including increases in vehicle surface heat flux and decreases in electron temperature. However, studying the ambipolar diffusion approximation in multi-species hypersonic flows (such as air flows) makes it challenging to isolate the origin of these differences.  Kinetic hypersonic flow simulations with electrostatic modeling are performed with argon to quantify the validity of the ambipolar diffusion approximation along a one-dimensional stagnation streamline. Three different plasma diffusion regimes are studied under two sets of rarefied freestream flow conditions. The approximation is evaluated in terms of predicting plasma density distributions, electron temperature, and stagnation point heat flux. New criteria are proposed for identification of plasma diffusion regimes in hypersonic flows and use of the ambipolar diffusion approximation. 
		
		
	\end{abstract}
	
	\maketitle
	
	\section{\label{sec:level1}Introduction and Background}
	
\begin{figure*}[tb]
		\centering 
		\includegraphics[width=0.65\textwidth]{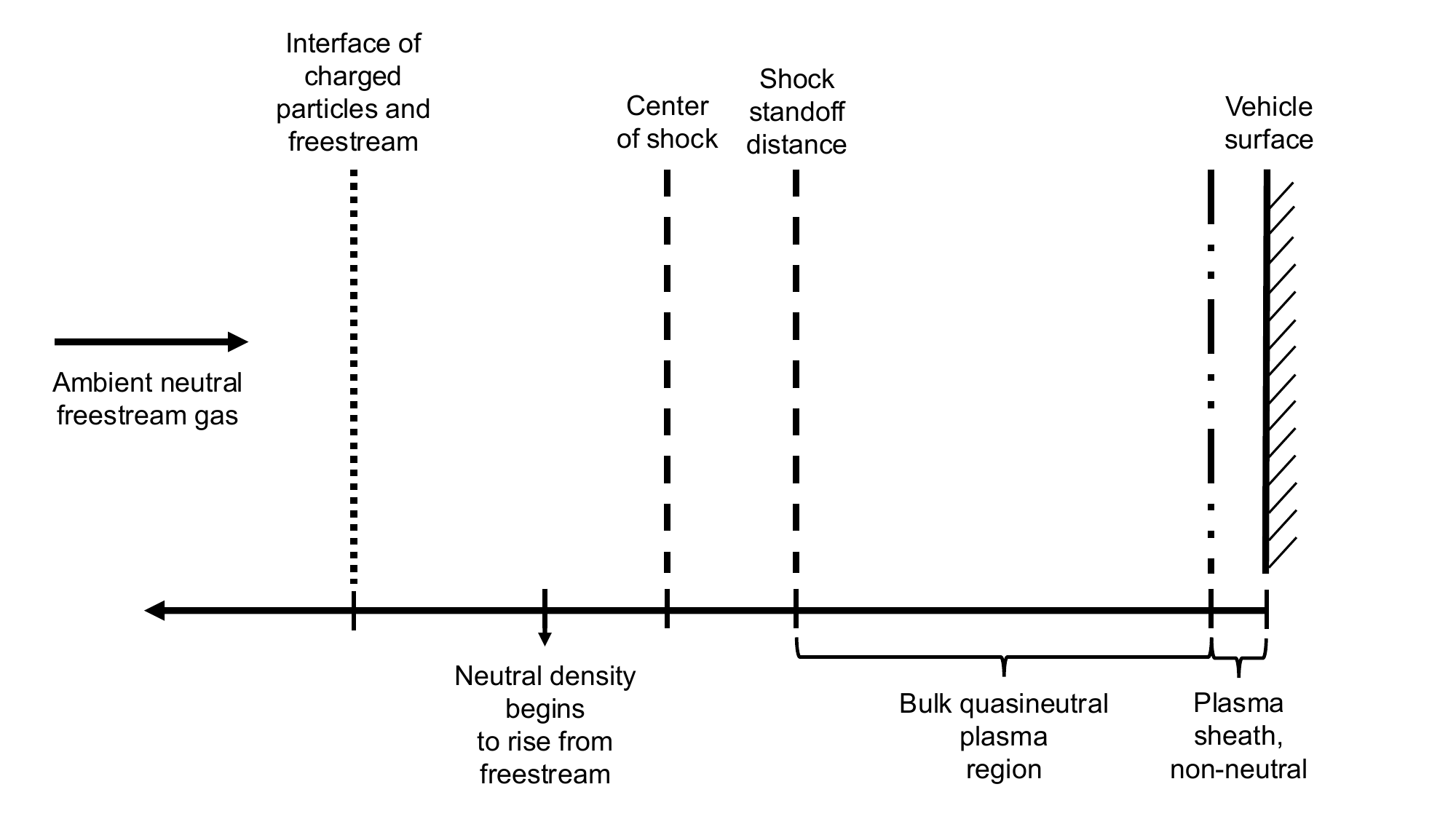}
		\caption{\label{fig:shock_schematic} Notional schematic of the plasma generated along a rarefied hypersonic stagnation streamline (not to scale).}
	\end{figure*}

	At hypersonic entry speeds, a spacecraft vehicle compresses the surrounding atmosphere, creating a bow shock that converts kinetic energy into thermal and internal energy. Depending on the flight conditions, the gas may ionize into a plasma. The conventional description for a partially ionized shock layer is depicted in Fig.~\ref{fig:shock_schematic}, in which the majority of the shock layer consists of a quasineutral plasma. Adjacent to the vehicle surface, a plasma sheath forms \footnote{In aerospace engineering, the term `plasma sheath' is sometimes used to refer to the entire partially ionized shock layer. We exclusively refer to the plasma sheath as the thin, non-charge neutral region at the vehicle surface.}, and upstream of the shock, the charged species densities gradually decrease towards zero. Accurate prediction of the hypersonic plasma environment via numerical modeling is vital to the development of advanced aerospace technologies, including radio communications blackout mitigation \cite{10.1063/5.0257810}, hypersonic vehicle remote observation \cite{hu_review_2020}, magnetohydrodynamic flow control \cite{pu_nonequilibrium_2026}, and electron transpiration cooling \cite{vatansever_numerical_2024,monroe_electron_2025}. At a more fundamental level, there is growing interest in high-fidelity characterization of non-equilibrium plasma chemistry and electron kinetics due to the role plasma species play in redistributing energy of the gas via excitation \cite{Petrov2024,petrova_quasi-one-dimensional_2025}. 
	
	Simulation of hypersonic flows with full electrostatic modeling, defined as modeling ions and electrons as independently moving species with electric-field dependent velocities and neglecting magnetic interactions, is computationally expensive. The challenge lies in resolving the vast differences in length and time scales between the hypersonic vehicle system and the plasma. Heavy ions and neutrals have velocities on the order of several thousand meters per second, whereas lightweight electrons have velocities 2 to 3 orders of magnitude larger. A typical characteristic length scale for a hypersonic system is on the order of 0.1 m, and the post-shock Debye length for a hypersonic plasma in air \cite{monroe_electron_2025} ranges between $10^{-5}$ and $10^{-7}$ m. This results in a stiff numerical system, in which extremely small integration time steps are required relative to the smoothness of the system, particularly when Gauss-law-based potential equations are used to solve for the self-induced electric fields \cite{parent_electron_2021}. 
	
	
	
	\begin{figure*}[tb]
		\centering 
		\includegraphics[width=0.65\linewidth]{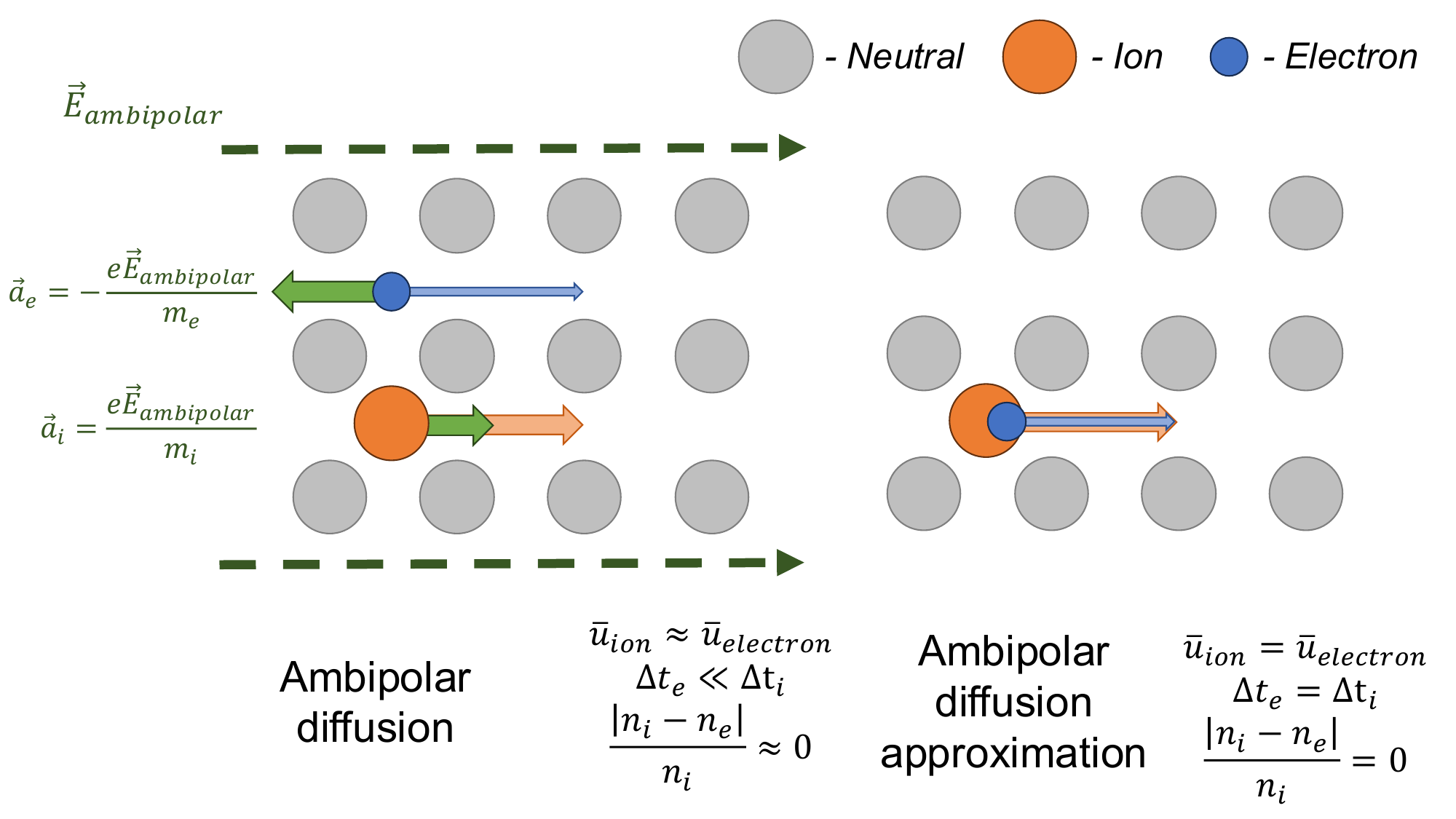}
		\caption{\label{fig:ambi_approx_diagram} (Not to scale) Left: Diagram of singly-charged plasma undergoing ambipolar diffusion. The ambipolar electric field is represented by dark green dashed lines. The acceleration vector for each charged particle is represented with a light green arrow. Right: Diagram of charged species transport with the ambipolar diffusion approximation and exact charge neutrality enforced. }
	\end{figure*}
	
	This motivates the near ubiquitous use of the ambipolar diffusion approximation (also known as the quasineutral assumption) \cite{Bird1985,Boyd1997}: a modeling approach that assumes the charge separation throughout the flowfield is sufficiently close to zero so as to justify neglecting charge separation-induced electrostatic effects. In kinetic simulation methods, the approximation is enforced by propagating ions and electrons with the same bulk velocity \cite{Boyd1997} or by assuming a number of electrons equal to the charge of an ion always remain sufficiently close to the ion such that the flow is exactly charge neutral \cite{Bird1985}. Because the ambipolar diffusion approximation does not compute electrostatic fields, resolution of the Debye length and fast electron motion is not required and significant computational expense is saved. However, the ambipolar diffusion approximation modeling technique is distinct from the plasma diffusion process by the same name, and the two should not be considered interchangeable. Diagrams of ambipolar diffusion and the ambipolar diffusion approximation are presented in Fig.~\ref{fig:ambi_approx_diagram}. The key difference is that with ambipolar diffusion, quasineutrality is maintained through a self-consistent ambipolar electric field, which can impact ion and electron properties via acceleration and deceleration. The ambipolar diffusion approximation maintains quasineutrality through enforced co-movement of ions and electrons. The purpose of this study is not to determine whether the ambipolar diffusion approximation captures ambipolar diffusion; this was investigated in Ref.~\onlinecite{petrusky_evaluation_2026}. Rather, the goal of this study is to broadly characterize the ability of the approximation to capture specific plasma macroscopic properties and physical processes in partially ionized rarefied hypersonic flows. 
	
	
	To date, comparatively few hypersonics modeling studies have attempted to formally quantify the tradeoffs of using the ambipolar diffusion approximation relative to simulations using some form of electrostatic modeling. Most studies conclude that when electrostatic modeling is used, non-negligible changes in charged species density distributions \cite{carlson_direct_1992,gallis_effect_1998,Farbar2010,Blanco2020,parent_electron_2021}, electron temperature profiles \cite{carlson_direct_1992,Blanco2020,parent_electron_2021}, and vehicle surface heat flux \cite{Farbar2010,Blanco2020} arise. Most recently, Ref.~\onlinecite{parent_electron_2021} defined the electron cooling effect: where in a hypersonic flow with very large electron density gradients, it is possible for the electron drift velocity component to point in the same direction as the self-induced ambipolar electric field. At steady state, the electron energy transport equation can be expressed as,
	
	\begin{equation}
		\sum_{i=1}^3 \rho_e \vec{u}_e \frac{\partial}{\partial x_i} h_e = \vec{E} \cdot \vec{J}_e + Q_{e,n} - Q_{EII}
		\label{eq:electron_energy_transport}
	\end{equation} 
	
	\noindent where $\rho_e$ is the mass density, $\vec{u}_e$ is the bulk velocity, $h_e$ is the enthalpy (proportional to temperature), $\vec{E}$ is the electric field, $\vec{J}_e \propto -\vec{u}_e$ is the electron current, $Q_{e,n}$ is the energy exchange between electrons and neutrals, and $Q_{EII}$ is the energy required for electron-impact ionization. Therefore, if $\vec{u}_e$ and $\vec{E}$ point in the same direction, electrons can lose energy to deceleration and $h_e$ may decrease, resulting in a lower electron temperature than would be obtained with the ambipolar diffusion approximation or quasineutral modeling. If $\vec{u}_e$ and $\vec{E}$ point in opposite directions, electron heating can occur instead. In Ref.~\onlinecite{parent_electron_2021}, this was reported to occur shortly downstream of the bow shock. Although this relation is derived from continuum equations, in kinetic flows, the relationships between terms can be used to broadly characterize electron energy transfer. Lower electron temperatures were also reported in Refs.~\onlinecite{carlson_direct_1992} and~\onlinecite{Blanco2020}. Considering electron-impact ionization (dictated by the electron temperature) is the primary source of plasma species at super-orbital Earth entry flow conditions \cite{lemal_prediction_2016}, the electron heating and cooling effects warrant further investigation. Additionally, when electrostatic modeling is used, Refs.~\onlinecite{Farbar2010} and~\onlinecite{Blanco2020} report increases in vehicle surface heat flux of 13-16\% and upwards of 31\%, respectively. These increases may be significant from a vehicle design perspective; however, higher-order quantities such as stagnation point heat flux are highly sensitive to freestream flow conditions and modeling choices. 
	
	A majority of the prior studies were limited to analyzing a few freestream flow conditions in air due to the computational expense. Moreover, it is unknown how well assessments of the ambipolar diffusion approximation in air translate to other atmospheres. This is in part because 11-species air (N$_2$, O$_2$, NO, N, O, each of their respective ions, and e$^-$) can experience upwards of 47 different chemical reactions. Thus, it is often unclear whether these differences originate directly from acceleration and deceleration of charged particles or from changes in the many coupled chemical reaction rates within the flow. The ambipolar diffusion approximation should also be analyzed in the context of the flowfield's plasma diffusion regime. There are three plasma diffusion regimes relevant to collisional partially ionized hypersonic flows: ambipolar diffusion, free diffusion, and transitional diffusion. For a continuum flow, ambipolar diffusion occurs if the partially ionized gas meets the formal definition of a plasma, such that collective plasma behavior dominates over hydrodynamic forces. This enables the formation of a self-induced electric field $\vec{E}_{ambipolar}$ that accelerates and decelerates charged particles such that they diffuse at the same rate and the plasma remains quasineutral (see Fig.~\ref{fig:ambi_approx_diagram}). Generally, for hypersonic plasmas where the Debye length is orders of magnitude shorter than the width of the shock layer, ambipolar diffusion is expected. If the partially ionized gas is primarily governed by hydrodynamic forces rather than electrostatic ones, the gas is in the free diffusion regime \cite{Phelps1990}. Characterized by large amounts of charge separation, the free diffusion regime can occur in regions of a hypersonic flow with large Debye lengths, such as in the wake region of a vehicle \cite{bartel_direct_1994}. The transitional diffusion regime is the intermediary stage between ambipolar and free diffusion, where the flow demonstrates some degree of collective plasma behavior, but not enough to preserve quasineutrality. Documented cases of the transitional flow regime have not been published in the hypersonics literature; one objective of this study is to distinguish transitional hypersonic plasma behavior from free and ambipolar diffusion plasma behavior. Development of robust criteria for identifying plasma diffusion regimes in arbitrary hypersonic flows would enable informed selection of plasma models and better understanding of hypersonic plasma behavior.
	
	Phelps developed criteria for characterizing the plasma diffusion regime in gas discharge vessels \cite{Phelps1990} based on the concept of the diffusion length \cite{McDaniel1964} $\Lambda$ and its ratios with the electron Debye length $\lambda_{D,e}$ and the ion mean free path $\lambda_{mfp,i}$ for collisions with neutrals. $\Lambda/\lambda_{mfp,i}$ describes how collisional the partially ionized gas is, and $\Lambda/\lambda_{D,e}$ describes the degree of collective plasma behavior of the gas (i.e., an assessment of quasineutrality). For both ratios, higher values indicate stronger behavior. If $\Lambda$ is broadly interpreted as the characteristic distance a charged particle travels before encountering a disturbance, then Phelps' criteria can be applied in contexts outside gas discharge vessels, such as unsteady expanding partially ionized gases \cite{petrusky_evaluation_2026}. Table \ref{tab:phelps_crit} lists the approximate ranges of $\Lambda/\lambda_{mfp,i}$ and $\Lambda/\lambda_{D,e}$ for each plasma diffusion regime. In Ref.~\onlinecite{petrusky_evaluation_2026}, the exact value of $\Lambda/\lambda_{D,e}$ for which a partially ionized gas entered the ambipolar diffusion regime ranged from 40 to 80 depending on the collision model used, the types of collisions simulated, and the value of $\Lambda/\lambda_{mfp,i}$. Thus, the lower limit of $\Lambda/\lambda_{D,e}$ for the ambipolar diffusion regime is highly flowfield and model dependent. 
	
		\begin{table}[tbh]
		\caption{\label{tab:phelps_crit} Limits of the free, transitional, and ambipolar diffusion regimes as defined by Refs.~\onlinecite{petrusky_evaluation_2026} and~\onlinecite{Phelps1990}. }
		\begin{ruledtabular}
			\begin{tabular}{lcc}
				Plasma diffusion regime & Range of $\Lambda/\lambda_{mfp,i}$ & Range of $\Lambda/\lambda_{D,e}$ \\
				\hline
				Free diffusion & $\Lambda/\lambda_{mfp,i} \gg 1.0$ & $\Lambda/\lambda_{D,e} \leq 1.0$  \\
				Transitional diffusion & $\Lambda/\lambda_{mfp,i} \gg 1.0$ & $1.0 < \Lambda/\lambda_{D,e} < 100$  \\
				Ambipolar diffusion & $\Lambda/\lambda_{mfp,i} \gg 1.0$ & $\Lambda/\lambda_{D,e} \gtrsim 100$ \\
			\end{tabular}
		\end{ruledtabular}
	\end{table}
	
	In this study, we perform a series of one-dimensional (1D) rarefied hypersonic stagnation streamline simulations in argon with a coupled Direct Simulation Monte Carlo (DSMC)\cite{BirdTB1994} - Particle-in-Cell (PIC) \cite{HockneyTB1981} solver in order to address two objectives: evaluating the efficacy of the ambipolar diffusion approximation in terms of reproducing results from an electrostatic simulation and determining whether the criteria for characterization of plasma diffusion regime used in Refs.~\onlinecite{petrusky_evaluation_2026} and~\onlinecite{Phelps1990} can be applied to hypersonic flows. Section \ref{sec:methodology} describes the numerical methods, physical models, and modeling assumptions used to conduct this study. Section \ref{sec:results} presents results and analyses. The main findings of the study are summarized and discussed in Section \ref{sec:summary}. We conclude in Section \ref{sec:conclusion}.
	
	\section{Methodology}
	\label{sec:methodology}
	
	\subsection{Numerical Methods}
	\label{sec:model}
	
	This study is performed with MONACO-PIC (MPIC): an in-house coupled DSMC-PIC code \cite{lipscomb_simulation_2026,stasiukevicius_framework_2026}. Advection of neutral particles and collision dynamics of all species are handled with DSMC, ionization reactions are handled with the Total Collision Energy model \cite{BirdTB1994}, and advection of charged species is handled with PIC. When the ambipolar diffusion approximation is invoked, `ambipolar particles' are created and assigned a set of ion velocity components and electron velocity components. During collisions, each ambipolar particle is split into an ion and an electron, each of which is considered for collision pairing independently. After collisions are performed, each ion is re-paired to an electron such that only ambipolar particles and neutrals remain during advection. A similar approach to the ambipolar diffusion approximation is used in the well-established open-source DSMC solver SPARTA\cite{SPARTA}. Only singly charged ions are modeled, therefore the pairings are always 1 to 1. The outputted electron number density is equal to the ion number density due to the enforcement of exact charge neutrality. The stored electron velocity components are used to output an `ambipolar electron temperature' separate from the ion temperature. 
	
	The presence of trace species in a DSMC-PIC flow can lead to an excess of non-trace particles resulting in an inefficient simulation. One way to alleviate this computational cost is to assign different weights to trace particles, known as variable-weight methods \cite{boyd_conservative_1996,rjasanow_stochastic_2005}. This work implements a semi-dynamic species weighting scheme (SWS). Details of how mass, momentum, and energy are conserved are included in Ref.~\onlinecite{boyd_conservative_1996}. On creation, each particle of species $i$ is assigned a weight $W_i = w_i f_{num}$, where $f_{num}$ is the global particle weight and $w_i$ is a fraction that modifies the global weight to give a species-specific weight. $w_i$ for ions and electrons are fixed, as they are always trace species throughout the flowfield and do not require dynamic weights. $w_i$ for individual neutrals may be reduced if they undergo electron-impact ionization or atom-atom ionization between neutrals of unlike weights. The weight maximum filter algorithm \cite{rjasanow_stochastic_2005} is used to pair particles for collisions, accounting for collisions between particles of unlike weights. 
	
	\subsection{Physical Models}
	
	The objective of this study is to characterize the impact of high-fidelity electrostatic modeling on prediction of rarefied hypersonic flows; therefore, the simulation setup should be simplified as much as possible while still being sufficiently representative of a hypersonic flight environment. As opposed to air, argon gas is selected for analysis. Modeling a monatomic gas circumvents consideration of rotational and vibrational degrees of freedom and limits the possible binary interactions between particles to momentum-exchange (MEX) collisions, ionization, and charge-exchange (CEX). Additionally, argon is a common choice of gas for aerospace plasma experiments due to its similar mass to air ($m_{air}/m_{Ar} = 0.723$). Since diffusion processes depend on particle mass, choosing a species similar in mass to air improves generalizability of results. Lastly, future missions to outer planets incentivize study of hypersonic plasma behavior in monatomic flows. 
	
	MEX collisions between neutrals and neutrals, neutrals and ions, and neutrals and electrons are modeled using the variable hard sphere (VHS) cross section, the Sakabe and Izawa CEX cross section  \cite{Sakabe1992}, and IST-Lisbon cross section data \cite{lxcat_lisbon,Alves2014}, respectively. Electron-impact ionization and atom-atom ionization are modeled: although these reaction rates may change with electrostatic modeling, they are required to generate a plasma within the domain and thus must be included. Excited states are not explicitly modeled; the selected ionization rates account for the `ladder-climbing' of excited states an argon atom will experience before ionizing-- characteristic of super-orbital hypersonic flows \cite{Annaloro2012}. The CEX reaction is not modeled in order to limit the total number of reactions present in the flow. Coulomb collisions are omitted because typical hypersonic flows in air are weakly ionized\cite{monroe_electron_2025} and charged-neutral collisions can dominate collisional transport. The derivation of ambipolar diffusion also assumes collisions with neutrals dominate over collisions between charged species\cite{ChenTB1984}. The strongly ionized argon conditions of this study represent an idealized test case, not ionization levels expected in realistic hypersonic air flows. Inclusion of Coulomb collisions under these conditions may overestimate their relative importance to the intended application. Electrons absorb at the wall, and ions recombine at the wall and are re-emitted as neutrals. Baseline parameters for all interactions are found in Appendix \ref{sec:appendix_argon}. To reduce the computational cost of resolving fast electron motion, the electron mass is artificially increased. Since choice of $m_e$ will lead to differences in ionization rates and density distributions throughout the flowfield, a sensitivity analysis is conducted in Section \ref{sec:mass_sens}. The results from Sections \ref{sec:free_diff} to \ref{sec:ambi_diff} use $m_i/m_e =$ 1,000.
	
	A 1D numerical model for simulating the stagnation streamline of a blunt body moving at a hypersonic speed for DSMC is used to simulate a rarefied hypersonic flowfield \cite{Bird1986, BirdTB1994}. The stagnation streamline is modeled as a constant streamwise area flow with an inflow boundary condition on one end and a wall boundary condition on the other. Particles enter the domain and reflect off the wall, resulting in an unsteady shock propagating upstream. To simulate a steady state flow, once the shock reaches a specified standoff distance, particles are selected and removed according to probabilistic criteria such that mass, momentum, and energy are conserved. If an ion is selected for removal, a random electron is also removed to conserve the original charge density of the cell. The model has successfully been used to study many different non-equilibrium phenomena in hypersonic flows, including electrostatic effects \cite{carlson_direct_1992,Farbar2010} and rarefied hypersonic flow behavior \cite{petrusky_discrete-velocity_2026}. Thus, the stagnation streamline model can reliably be used to characterize plasma behavior in hypersonic flows independent of perpendicular transport. 
	
	In realistic hypersonic flows, freestream conditions such as altitude and velocity impact shock layer width, surface temperature, plasma density, and plasma diffusion regime. In order to investigate a broad range of plasma diffusion conditions across a consistent hypersonic flowfield setup, a method of artificially reducing the freestream densities of the flowfield is used \cite{Farbar2010,kawamura_double_2009}. For each of the baseline flow cases described in Table~\ref{tab:default_param}, the freestream number density is reduced by the amount required to bring the maximum plasma density down to a specified magnitude; this factor is labeled $F_r$. The collision cross sections are then multiplied by $F_r$ to yield the same collision mean free path and reaction rates as the baseline flow. This approach ensures the distributions of axial velocity and temperature, ionization fraction profile, and the width of the shock layer remain similar to the baseline flow case within 3.95\% root-mean-square error (RMSE) (when modeled with the ambipolar diffusion approximation). The stagnation point heat flux is also scaled by a factor of $F_r$ to facilitate comparison with realistic flow conditions. The heat fluxes from the baseline simulations and the reduced density simulations agree within 3.15\% on a species-by-species basis and within 1.72\% for total stagnation point heat flux. This provides confidence that the artificial reduced density method is correctly scaling quantities of interest, facilitating qualitative comparison with other studies in the literature. However, because the ionization fractions are preserved from the baseline flow conditions, the flow cases do not represent the ionization levels expected in typical hypersonic air or argon flows.
	
	\subsection{Computational Domain Setup}
	
	The flow conditions are loosely based on the re-entry trajectory of the Stardust Sample Return Capsule \cite{olynick_aerothermodynamics_1999,boyd_modeling_2010}. Two points along the trajectory are chosen for analysis: 60 km at Mach 35 and 81 km at Mach 39. At each altitude, the shock standoff distance for the stagnation streamline model is selected to yield a shock layer similar in width to axisymmetric Stardust simulation data in air. Each flow condition is simulated with three different values of $F_r$ to capture each of the plasma diffusion regimes of interest. Baseline parameters for the flow conditions are listed in Table \ref{tab:default_param}, and specific case parameters are listed in Tables \ref{tab:specific_param_60km} and \ref{tab:specific_param_81km}. The global particle weight is chosen to yield at least 20 neutral particles per cell in the freestream. The species weight modifiers for Ar$^+$ and e$^-$ are selected to yield at least 100 particles per cell in the bulk plasma region. Each simulation is run to steady state, and the final macroscopic property profiles are time averaged over $10^6$ time steps in the steady state period. 
	
	Configuration space grids are uniformly generated such that the local mean free path and \textit{minimum} electron Debye length of the entire flowfield are resolved with at least 2 points. A grid sensitivity analysis is run using the Mach 39 flow at 81 km with $F_r = 10^9$ flow case, and confirms that changing the grid resolution by a factor of 0.5 and a factor of 2 ensures the number density, axial velocity, and temperature profiles agree within 1.45\% and 2.50\% RMSE for ions and electrons, respectively. The heat fluxes agree within 3.22\% on a species-by-species basis, with the exception of the electrons whose heat fluxes agree within 5.16\% with increasing grid resolution. This aligns with expectations: since the uniform grid cell size is selected based on the minimum Debye length, a majority of the computational domain will have a converged grid cell size and the RMSE values for the entire macroscopic property profiles are small. Near the wall, charged particle motion may be more sensitive to changes in grid resolution, leading to larger differences in the electron heat flux. Since detailed study of hypersonic plasma sheath dynamics is beyond the scope of this work and electrons contribute less than 1\% of the total heat flux in all cases, the grid resolution applied throughout this work is considered sufficient. 
	
	\begin{table}[tbh]
		\caption{\label{tab:default_param} Initialization parameters and flowfield properties for each flow condition of this study. }
		\begin{ruledtabular}
			\begin{tabular}{lcc}
				Parameter & 60 km & 81 km \\
				\hline
				Freestream number density, molecules/m$^3$ & 4.87 $\times 10^{21}$ & 2.64 $\times 10^{20}$ \\
				Freestream temperature, K &  238 & 222 \\
				Freestream velocity, m/s & 10,350 & 10,810 \\
				Freestream Mach number & 35 & 39 \\
				Freestream mean free path, m & 2.55 $\times 10^{-4}$ & 4.59 $\times 10^{-3}$ \\ 
				Wall temperature, K& 3750 & 2000 \\
				Time step, s & $10^{-10}$ & $10^{-9}$  \\
				Shock standoff distance from wall, m & 0.00300 & 0.0243\\
			\end{tabular}
		\end{ruledtabular}
	\end{table}
	
	\begin{table}[tbh]
		\caption{\label{tab:specific_param_60km} Initialization parameters for each case simulated with the Mach 35 at 60 km flow condition. }
		\begin{ruledtabular}
			\begin{tabular}{lccc}
				Parameter & Free & Transitional & Ambipolar\\
				\hline
				Domain length, m & 0.03 & 0.03 & 0.03  \\
				Uniform grid width $\Delta x$, m &2.22 $\times 10^{-5}$ & 2.22 $\times 10^{-5}$ & 7.94 $\times 10^{-6}$ \\
				Points per minimum $\lambda_{D,e}$ &$\geq$ 102 & $\geq$ 7.59 & $\geq$ 2.16 \\
				Reduced density factor, $F_r$ & $10^{10}$ & $10^{8}$ & $10^6$  \\
				Global weight & 5.41 $\times 10^{5}$ &  5.41 $\times 10^7$ & 3.87 $\times 10^9$\\ 
				Plasma weight modifier & 0.05 & 0.05 & 0.05 \\
			\end{tabular}
		\end{ruledtabular}
	\end{table}
	
	\begin{table}[tbh]
		\caption{\label{tab:specific_param_81km} Initialization parameters for each case simulated with the Mach 39 at 81 km flow condition.}
		\begin{ruledtabular}
			\begin{tabular}{lccc}
				Parameter & Free & Transitional & Ambipolar \\
				\hline
				Domain length, m & 0.30 & 0.30 & 0.2  \\
				Uniform grid width $\Delta x$, m & 4.00 $\times 10^{-4}$ & 4.00 $\times 10^{-4}$ & 1.26 $\times 10^{-4}$ \\
				Points per minimum $\lambda_{D,e}$ & $\geq$ 21.9 & $\geq 5.72$ & $\geq$ 2.81\\
				Reduced density factor, $F_r$ & $10^{10}$ & $10^{9}$& $10^7$  \\
				Global weight & 5.28$\times 10^5$ & 5.28 $\times 10^6$ & 1.66 $\times 10^8$\\ 
				Plasma weight modifier & 0.05 & 0.05 & 0.10 \\
			\end{tabular}
		\end{ruledtabular}
	\end{table}
	
	To solve Poisson's equation for electrostatics for the electric potential, a zero-gradient Neumann boundary condition ($\phi '(x = -L) = 0$ V/m) is applied to the inflow boundary and a zero-value Dirichlet boundary condition ($\phi(x = 0) = 0$ V) is applied to the wall boundary.  A parametric survey is performed for the 81 km flow case with $F_r = 10^9$ where the total length of the computational domain $L$ is varied between $L = 0.2$ and $L = 0.5$ m. For comparison, the width of the shock layer is roughly 0.03 m. The electric field profiles agree within a maximum of 3.22\% RMSE in the shock layer region between all flow cases, and the charged species density distributions and electron temperature profiles agree within 0.948\% RMSE in the shock layer. Therefore, converged Poisson solutions are achieved with the domain lengths selected.
	
	\section{Simulation Results}
	\label{sec:results}
	
	Throughout this section, each case is assessed in terms of normalized charge separation ($= |n_i - n_e|/n_i$) expressed as a percent, the ratio of the diffusion length to the electron Debye length $\Lambda/\lambda_{D,e}$, and, from the third criterion of the definition of a plasma, $\omega \tau = \omega_{p,e} \tau_{e,n}$, where $\omega_{p,e}$ is the characteristic electron oscillatory frequency and $\tau_{e,n}$ is the mean collision time of electrons with neutrals. Electron temperature and stagnation point heat flux \cite{ko_radiative_2026} are also discussed due to their importance in vehicle design.
	
	Successful application of the criteria for identifying charged species diffusion regime outlined in Refs.~\onlinecite{petrusky_evaluation_2026} and~\onlinecite{Phelps1990} requires proper selection of a diffusion length. For a stagnation streamline, three key flowfield characteristics must be captured. The first is the plasma sheath: $\Lambda/\lambda_{D,e}$ should be of the order of magnitude of 1 or less within the sheath, as ambipolar diffusion is not expected when the plasma is not quasineutral. For flows in the ambipolar diffusion regime, the profile of $\Lambda/\lambda_{D,e}$ within the shock layer should be on the order of 10 or more. Finally, as one moves upstream the charged species densities decrease towards zero, therefore $\lambda_{D,e}$ should increase and $\Lambda/\lambda_{D,e}$ should decrease. These constraints rule out any constant values of $\Lambda$, such as setting $\Lambda$ equal to the length of the computational domain or the width of the shock layer, as they would not account for these ranges of length scales. We define the local diffusion length $\Lambda_h$ for a 1D parallel plane within a hypersonic flowfield as
	
	\begin{equation}
		\label{eq:hypersonic_diff_length}
		\Lambda_h(x) = {x - x_w}
	\end{equation}
	
	\noindent where for $\Lambda/\lambda_{D,e}$ calculated at location $x$, $x - x_w$ is the shortest distance to the wall or vehicle surface. Within the bulk plasma region where diffusion is expected to be ambipolar, $\Lambda_h$ is similar in magnitude to the shock layer width and thus is representative of the typical distance a charged species will travel to the wall. Since plasma sheaths have a width on the order of several $\lambda_{D,e}$, $\Lambda_h/\lambda_{D,e}$ will have an order of magnitude of 1.0 or less within the sheath. Demonstration of the effectiveness of Eq.~(\ref{eq:hypersonic_diff_length}) in predicting plasma diffusion regime is provided throughout this section.
	
	The error due to numerical statistical noise for the stagnation point heat flux is on the order of 0.100\%, and the largest reported error from the grid-convergence study is 5.16\%. Subsequently, differences in stagnation point heat flux below 5\% are considered negligible in the following analysis.
	
	\subsection{Free Diffusion}
	\label{sec:free_diff}
	
	\begin{figure*}[tb]
		\centering 
		\begin{subfigure}{0.49\linewidth}
			\includegraphics[width=\linewidth]{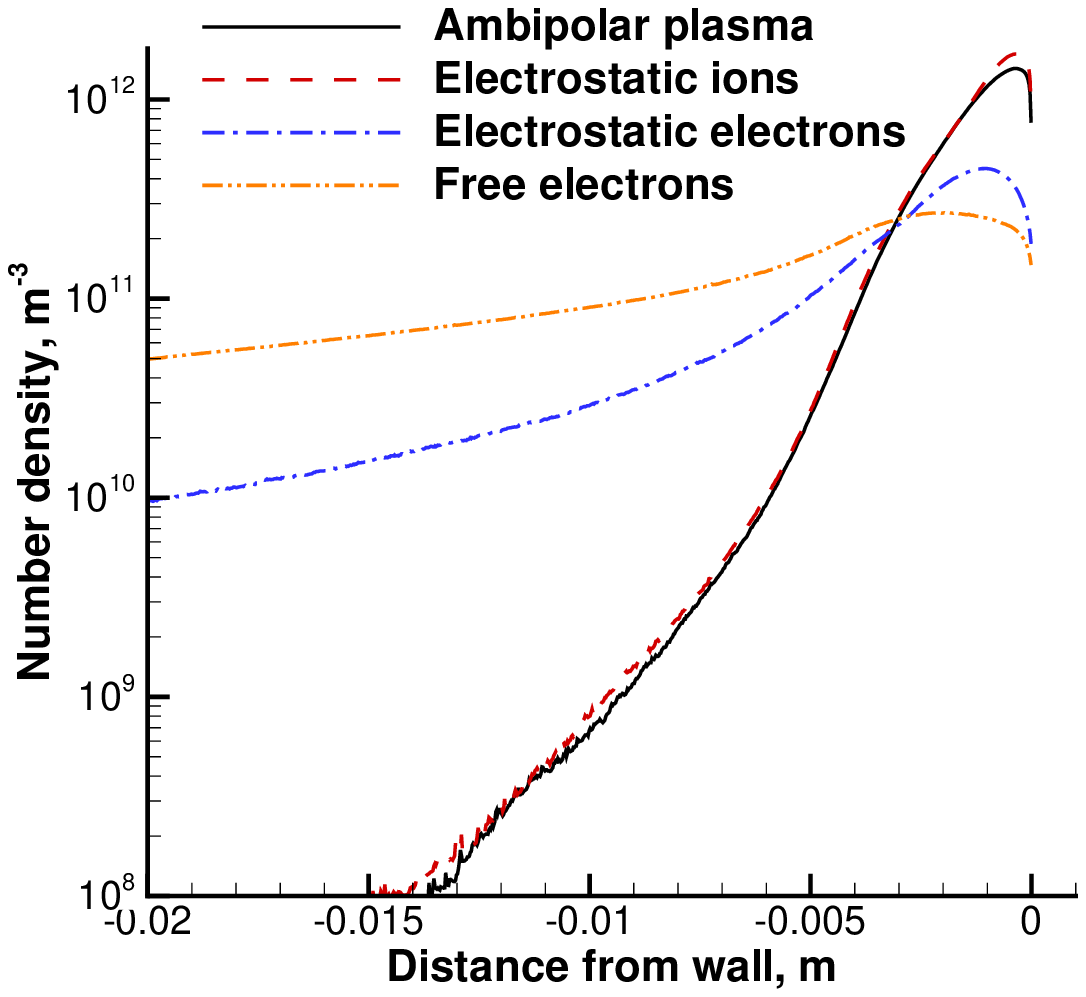}
			\caption{Charged species densities.}
			\label{fig:60km_Fr=1e10_dens}
		\end{subfigure}
		\hfill
		\begin{subfigure}{0.49\linewidth}
			\includegraphics[width=\linewidth]{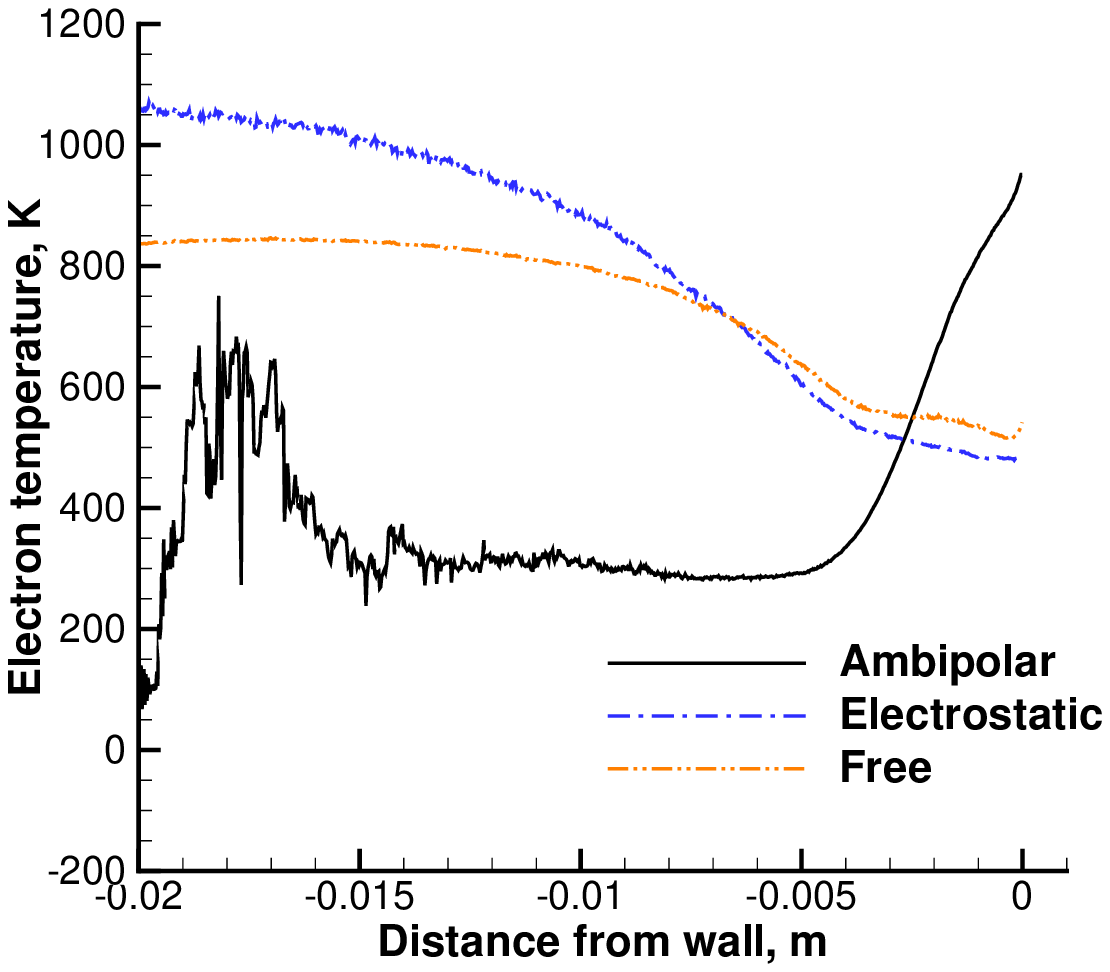}
			\caption{Electron temperature.}
			\label{fig:60km_Fr=1e10_elec_temp}
		\end{subfigure}
		\caption{Stagnation streamline flowfield under free plasma diffusion conditions with the ambipolar diffusion approximation, with full electrostatic modeling, and with the free diffusion approximation (Mach 35 flow, 60 km,  $F_r = 10^{10}$).}
		\label{fig:60km_Fr=1e10}
	\end{figure*} 
	
		\begin{figure}[tb]
		\centering 
		\includegraphics[width=0.49\linewidth]{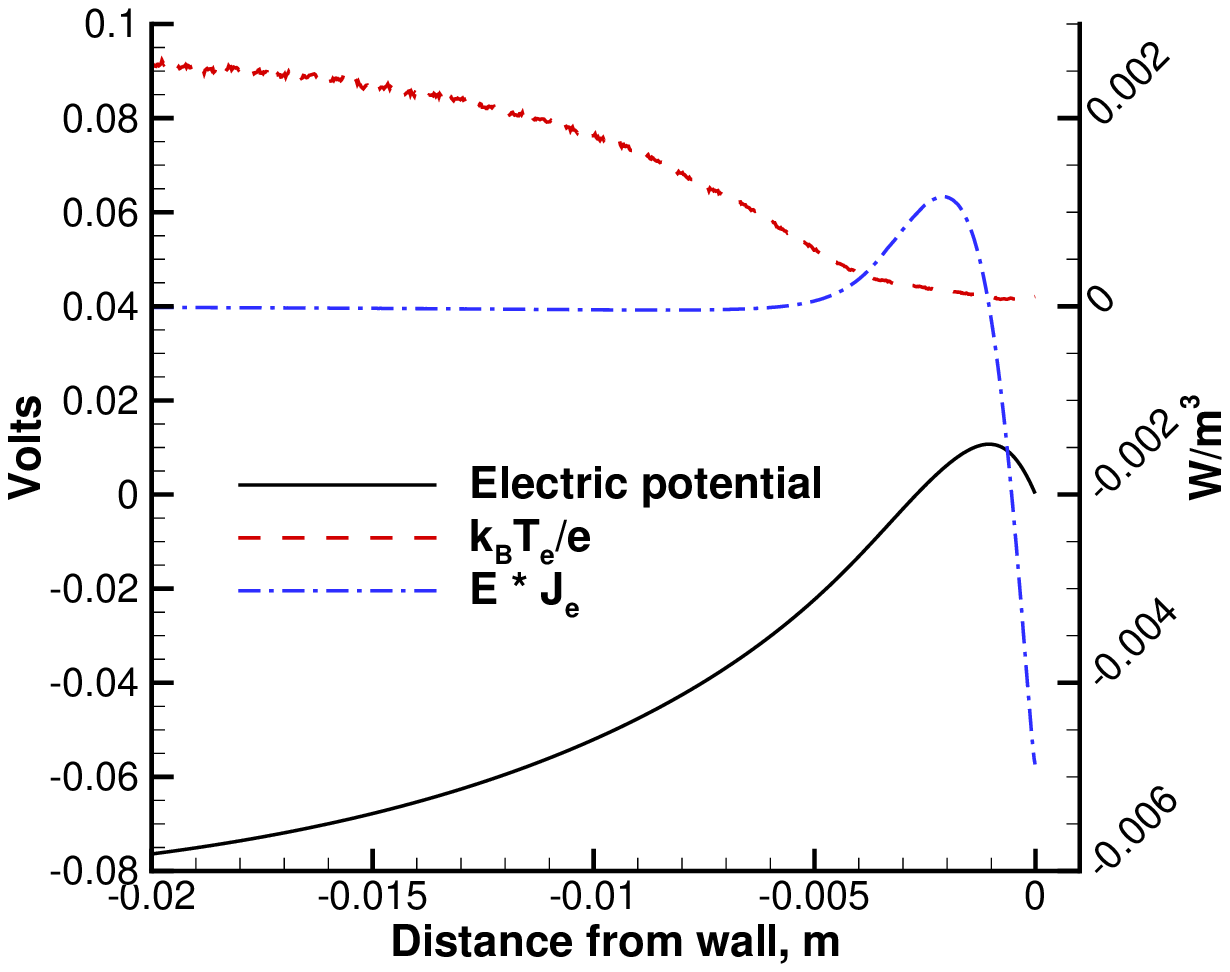}
		\caption{(Left axis) Electric potential and electron thermal energy per unit charge; (Right axis) Dot product of the electric field and the electron current density (Mach 35 flow, 60 km,  $F_r = 10^{10}$).}
		\label{fig:60km_Fr=1e10_other}
	\end{figure} 
	
	\begin{figure*}[tb]
		\centering
		\begin{subfigure}{0.32\linewidth}
			\includegraphics[width=\linewidth]{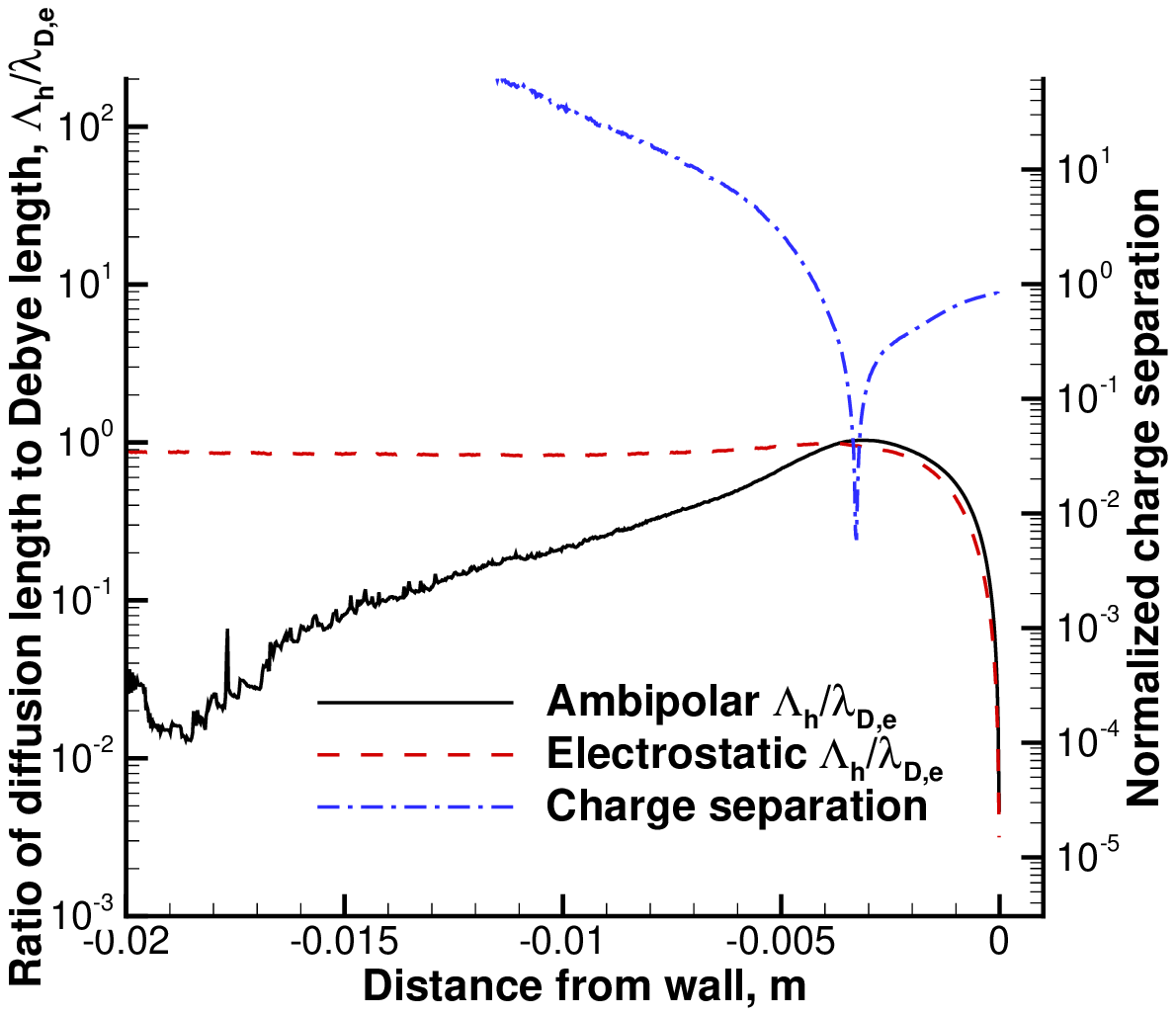}
			\caption{Free diffusion case ($F_r = 10^{10}$).}
			\label{fig:60km_Fr=1e10_diff_length}
		\end{subfigure}
		\hfill
		\begin{subfigure}{0.32\linewidth}
			\includegraphics[width=\linewidth]{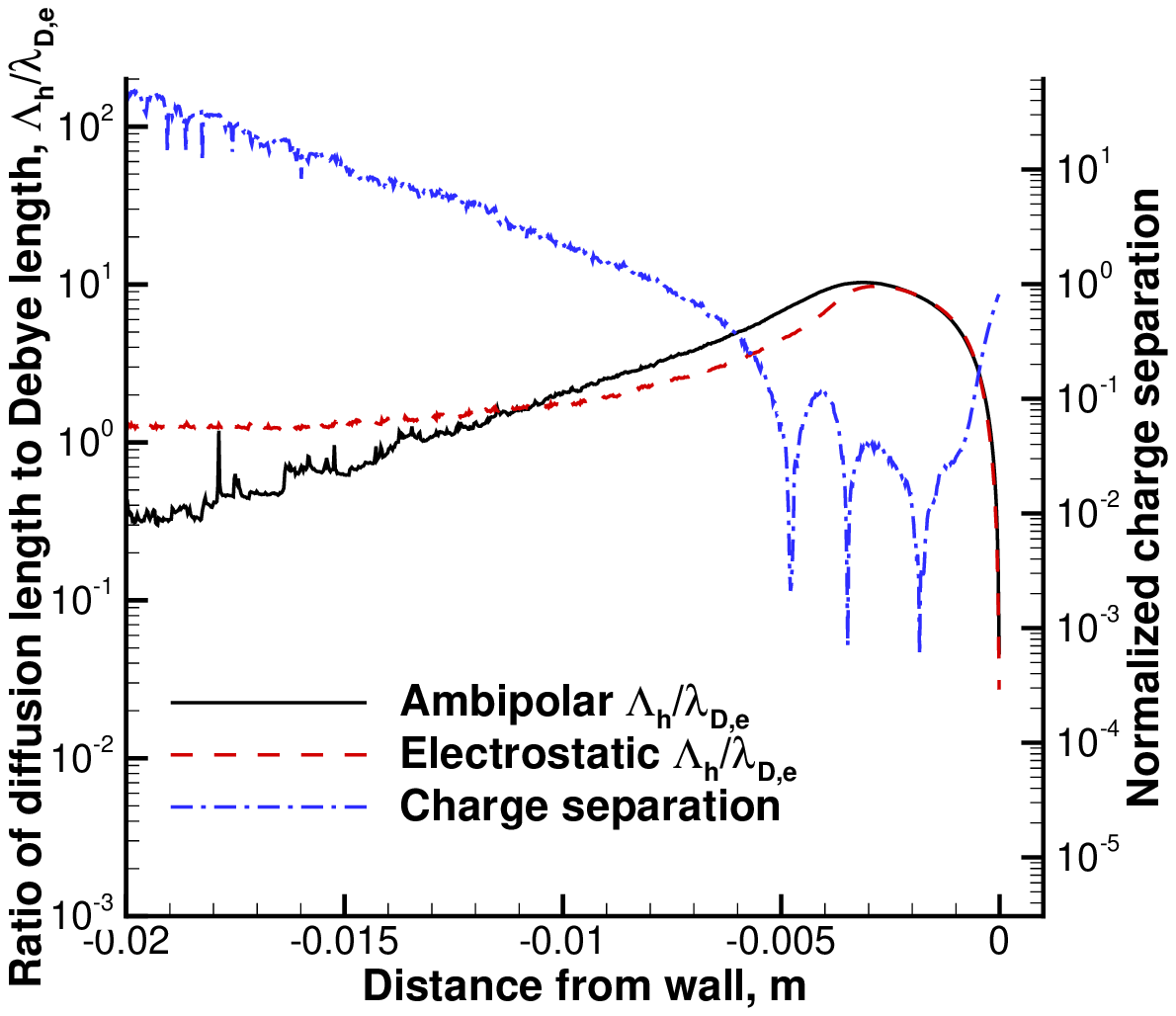}
			\caption{Transitional diffusion case ($F_r = 10^{8}$).}
			\label{fig:60km_Fr=1e8_diff_length}
		\end{subfigure}
		\hfill
		\begin{subfigure}{0.32\linewidth}
			\includegraphics[width=\linewidth]{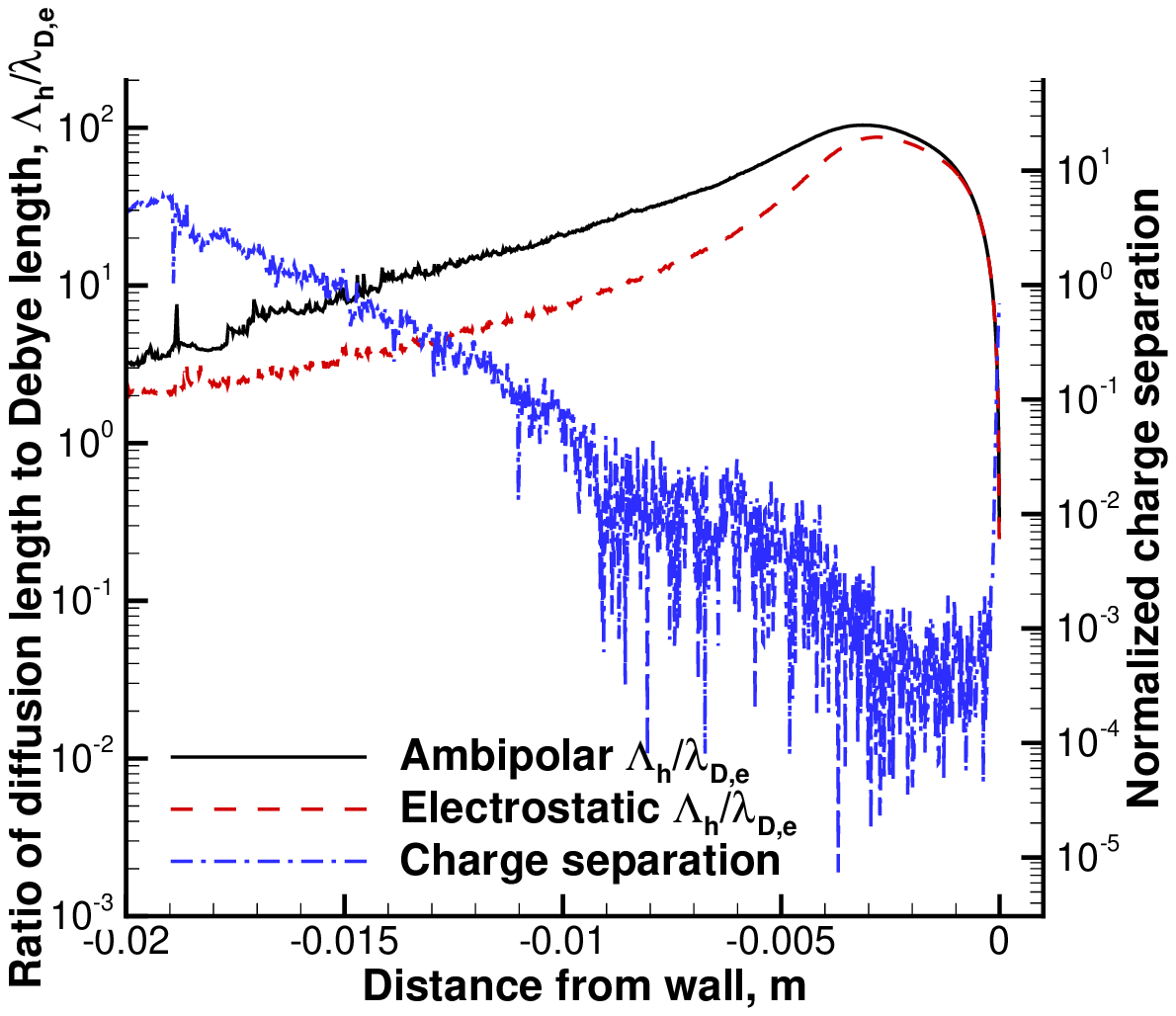}
			\caption{Ambipolar diffusion case ($F_r = 10^{6}$).}
			\label{fig:60km_Fr=1e6_diff_length}
		\end{subfigure}
		\caption{ Ratio of hypersonic diffusion length $\Lambda_h$ to electron Debye length $\lambda_{D,e}$ along a stagnation streamline with the ambipolar diffusion approximation and with full electrostatic modeling (Mach 35 flow, 60 km).}
		\label{fig:60km_diff_length}
	\end{figure*}
	
	We define the free plasma diffusion regime in hypersonic flows as having no portion of the shock layer undergo ambipolar diffusion, evidenced by the values of $\omega_{p,e}\tau_{e,n} < 1.0$, $\Lambda_h/\lambda_{D,e} < 1.0$, and charge separation above 10\% for a majority of the shock layer. Both cases in this section use $F_r = 10^{10}$. Figure~\ref{fig:60km_Fr=1e10} presents flowfield results with the ambipolar diffusion approximation and with full electrostatic modeling for the Mach 35 flow at 60 km. Select results from the free diffusion approximation solution (in which ions and electrons are modeled independently, but the Poisson equation is not solved) are also shown. Charge separation exceeds 10\% throughout the domain except near the intersection of the ion and electron density profiles from $-0.00344 < x < -0.00307$ m. This occurs because within the shock layer, the electron thermal energy per unit charge is roughly 4 times that of the electric potential (Fig.~\ref{fig:60km_Fr=1e10_other}), thus the electrons have sufficient energy to escape the self-induced electric fields and diffuse upstream. The partially ionized gas violates the third criterion for the definition of a plasma: $\omega_{p,e} \tau_{e,n} = 0.590$ at the point of peak electrostatic electron density ($x = -0.00107$ m) and $\omega_{p,e} \tau_{e,n} = 0.332$ at the point of peak ambipolar plasma density ($x = -0.000295$ m). The profiles of $\Lambda_h/\lambda_{D,e}$ from the electrostatic and ambipolar diffusion approximation solutions (Fig.~\ref{fig:60km_Fr=1e10_diff_length}) also confirm the presence of free diffusion, with $\Lambda_h/\lambda_{D,e} < 0.982$ for the electrostatic profile and $\Lambda_h/\lambda_{D,e} < 1.03$ for the ambipolar diffusion approximation profile. These results indicate the partially ionized gas in the free diffusion regime, consistent with the criteria outlined in Refs.~\onlinecite{petrusky_evaluation_2026} and~\onlinecite{Phelps1990}. If instead, the shock standoff distance is used ($L = 0.00300$ m), $L/\lambda_{D,e} < 1.32$, but the profile does not capture the sharp decrease at the wall due to the fixed value of $L$.
	
	At the shock standoff distance, the ambipolar diffusion approximation solution's electron temperature is 453 K and increases to 947 K at the wall. Conversely, the electrostatic solution's electron temperature profile is 519 K at the shock standoff distance and decreases to 487 K, with a flatter shape overall. The free diffusion approximation demonstrates a similar trend. Both the electrostatic and free diffusion approximation solutions increase in temperature as one moves upstream. These differences with the ambipolar diffusion approximation solution can be explained through Eq.~\ref{eq:electron_energy_transport}. Within the shock layer, the electron-impact ionization rates for all solutions are negligible since all electron temperature profiles are below 1,200 K (see Table \ref{tab:argon_ionization_param}); thus $Q_{EII}$ is negligible. The neutral temperature exceeds the electron temperature by 2 orders of magnitude throughout the shock layer; thus energy transfer occurs from neutrals to electrons and $Q_{e,n}$ will be positive. Since the ambipolar diffusion and free diffusion approximation solutions have no electric field, $\vec{E} \cdot \vec{J}_e = 0$. Therefore, the differences in the free and ambipolar diffusion approximation solutions can only arise from differences in $Q_{e,n}$. From Ref.~\onlinecite{parent_electron_2021}, $Q_{e,n} \propto n_e$; therefore $Q_{e,n}$ is smaller in magnitude within the shock layer for the free and electrostatic solutions compared to the ambipolar solution, leading to lower electron temperatures. Upstream, each electron number density profile is higher than the ambipolar plasma density profile, leading to higher electron temperatures. Between the electrostatic and free diffusion approximation solutions, the neutral density and temperature profiles agree within 0.0533\% and 0.333\% RMSE, respectively. Therefore, $Q_{e,n}$ is expected to be similar between solutions, resulting in their similar electron temperature profiles.
		
	Despite being in the free diffusion regime, the self-induced electric field still has an impact on flowfield properties. The free diffusion approximation underpredicts the peak electron number density by 43.6\% relative to the electrostatic solution. Diffusion rates are expected to be similar between the models due to their similar neutral number density  and electron temperature profiles. Therefore, this difference must arise from the self-induced electric field. From $-0.00581 < x < -0.00103$ m (Fig.~\ref{fig:60km_Fr=1e10_other}), $\vec{E} \cdot \vec{J}_e$ is positive, enabling more electrons to remain within the shock layer than the free diffusion approximation. $\vec{E} \cdot \vec{J}_e < 0$ near the wall, but the resulting electron cooling contributions are small relative to $Q_{e,n}$ due to the small magnitude of the electric field.
	
	
	The influence of independent electron motion on flowfield properties is also evident from the electrostatic convective heat flux, which increases by 94.8\% compared to the ambipolar diffusion approximation solution (Table \ref{tab:ch5_free_diff_heat_flux}). This appears counter-intuitive, since the electrostatic electrons have a lower temperature and density throughout the shock layer. When electrons are allowed to move independently, they will diffuse to the wall faster than ions due to their lightweight mass and the large density gradient that forms. This leads to a larger number flux of electrons at the wall. Although the electron convective heat flux increases with electrostatic modeling, it contributes to less than 1\% of the total stagnation point heat flux due to the electrons' lightweight mass. The electrostatic solution reports a decrease in total heat flux of $-7.44$\% relative to the ambipolar diffusion approximation. Since the change in contribution to heat flux from plasma species is only $+2.20$\%, the difference in total heat flux must be dominated by the decrease in neutral convective heat flux of $-10.4$\%. Evidently, independent mobility of electrons enables more transfer of energy from neutrals to electrons, which the electrons either carry upstream or to the wall. The ion convective heat flux also decreases by $-8.30\%$; since the electric field has little impact on the ion density profile, this must result from ions transferring energy to neutrals via collisions, then neutrals transferring that energy to electrons via collisions. The free diffusion approximation also observes similar trends in heat fluxes when compared to the ambipolar diffusion approximation, with a decrease in neutral convective heat flux of $-13.0\%$, a decrease in ion convective heat flux of $-10.8\%$, an increase in electron heat flux of $+107\%$, and decrease in total heat flux of $-6.45\%$. Therefore, the trends in convective heat fluxes are consistent between the electrostatic and free diffusion approximation solutions.

		\begin{figure*}[tb]
		\centering 
		\begin{subfigure}{0.49\linewidth}
			\includegraphics[width=\linewidth]{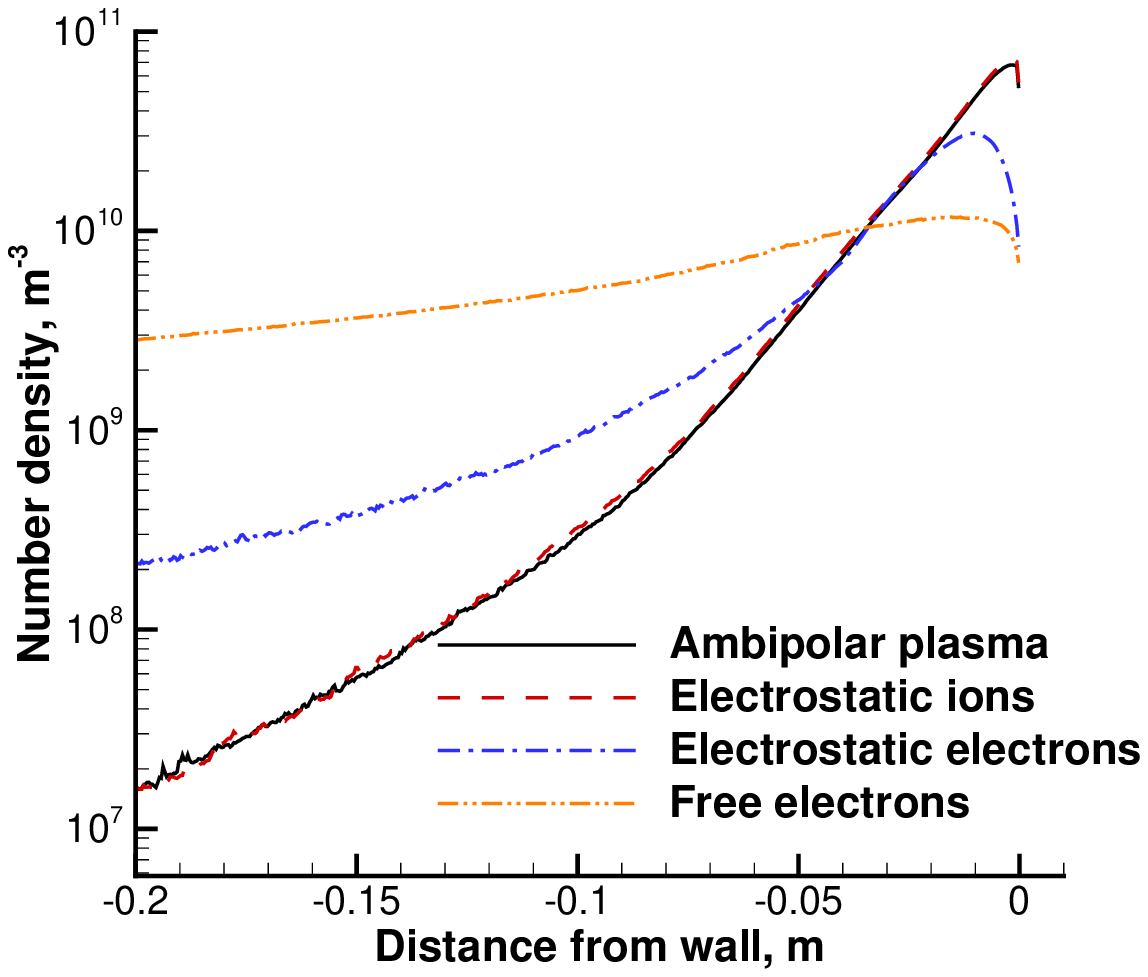}
			\caption{Charged species densities.}
			\label{fig:81km_Fr=1e10_dens}
		\end{subfigure}
		\begin{subfigure}{0.49\linewidth}
			\includegraphics[width=\linewidth]{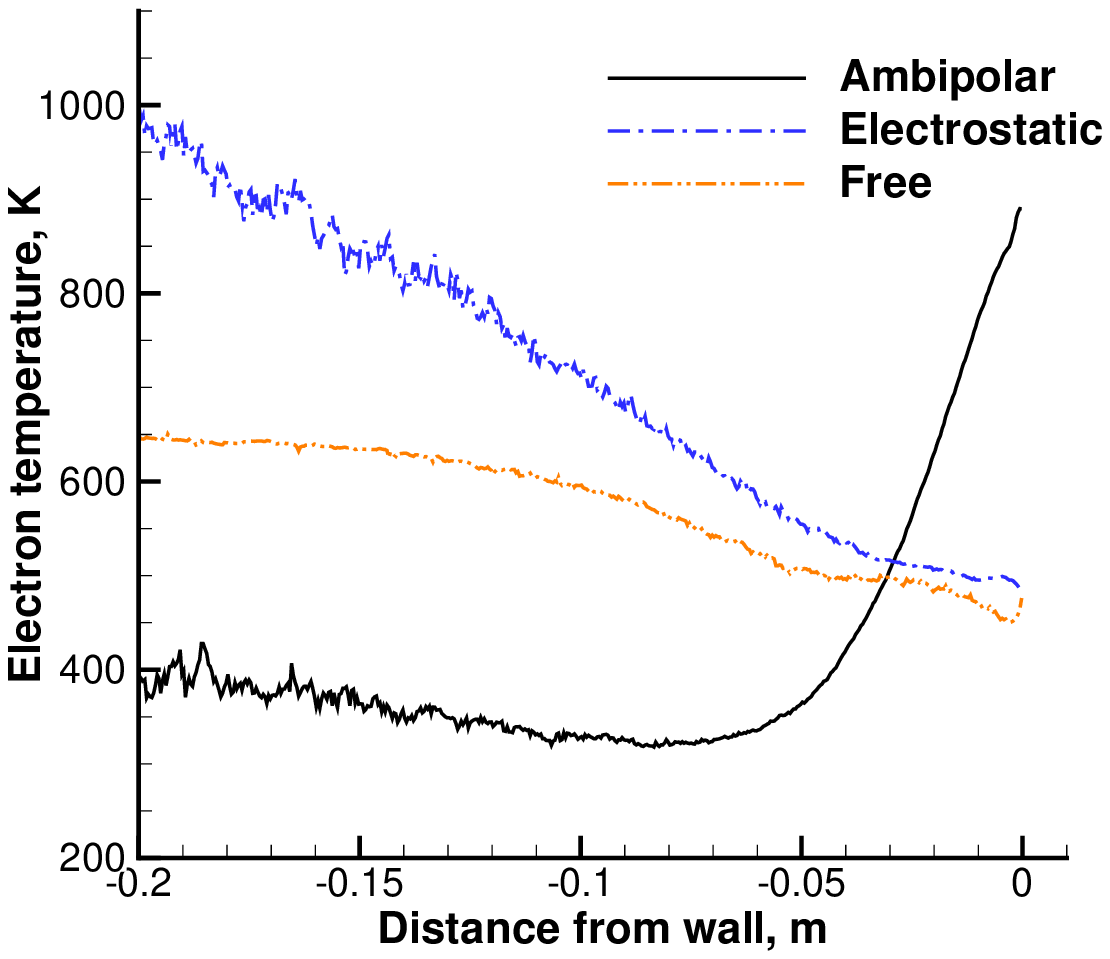}
			\caption{Electron temperature.}
			\label{fig:81km_Fr=1e10_elec_temp}
		\end{subfigure}
		\caption{Stagnation streamline flowfield under free plasma diffusion conditions with the ambipolar diffusion approximation, with full electrostatic modeling, and with the free diffusion approximation (Mach 39 flow, 81 km,  $F_r = 10^{10}$).}
		\label{fig:81km_Fr=1e10}
	\end{figure*} 
	
\begin{figure*}[tb]
		\centering
		\begin{subfigure}{0.32\linewidth}
			\includegraphics[width=\linewidth]{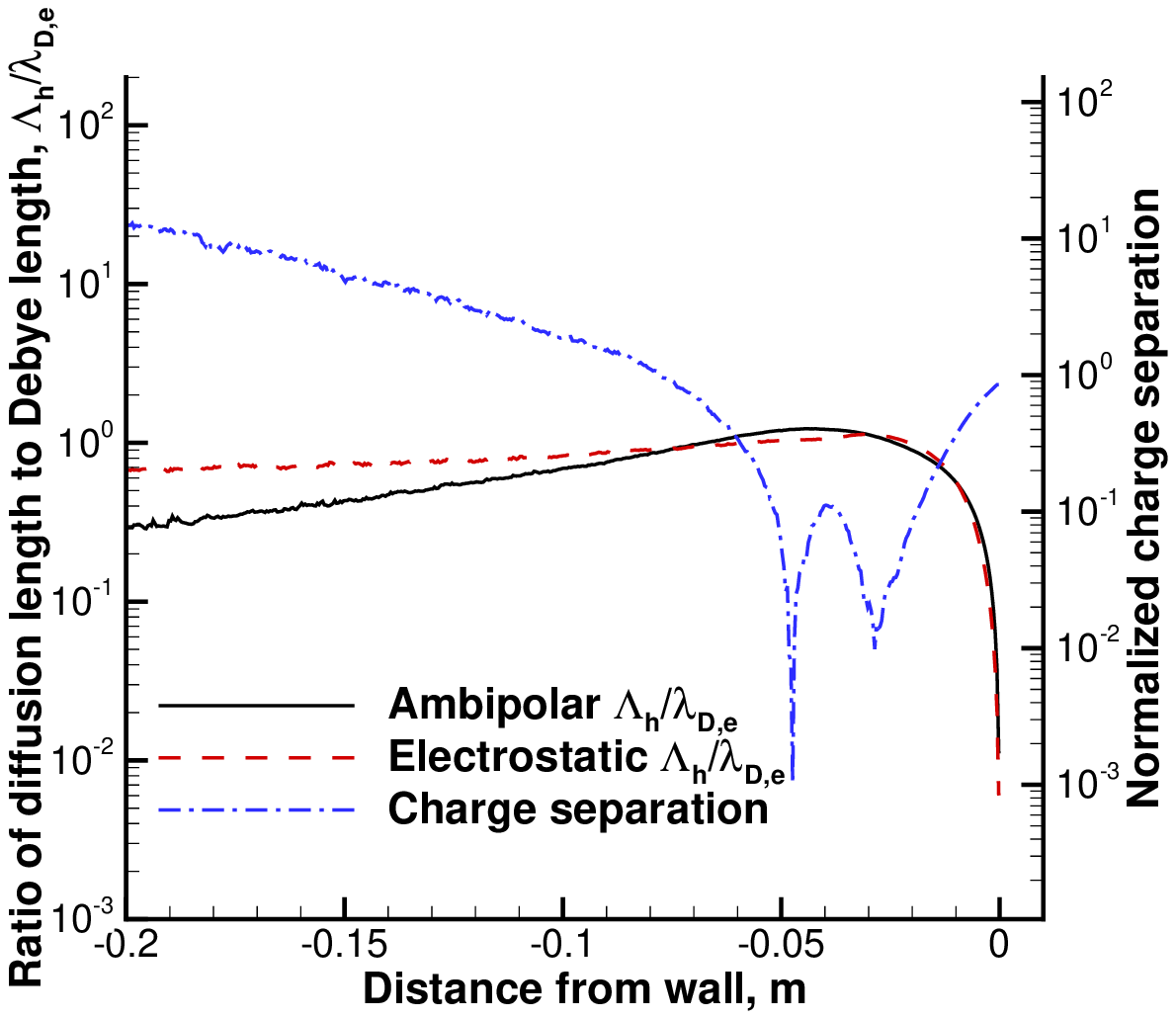}
			\caption{Free diffusion case ($F_r = 10^{10}$).}
			\label{fig:81km_Fr=1e10_diff_length}
		\end{subfigure}
		\begin{subfigure}{0.32\linewidth}
			\includegraphics[width=\linewidth]{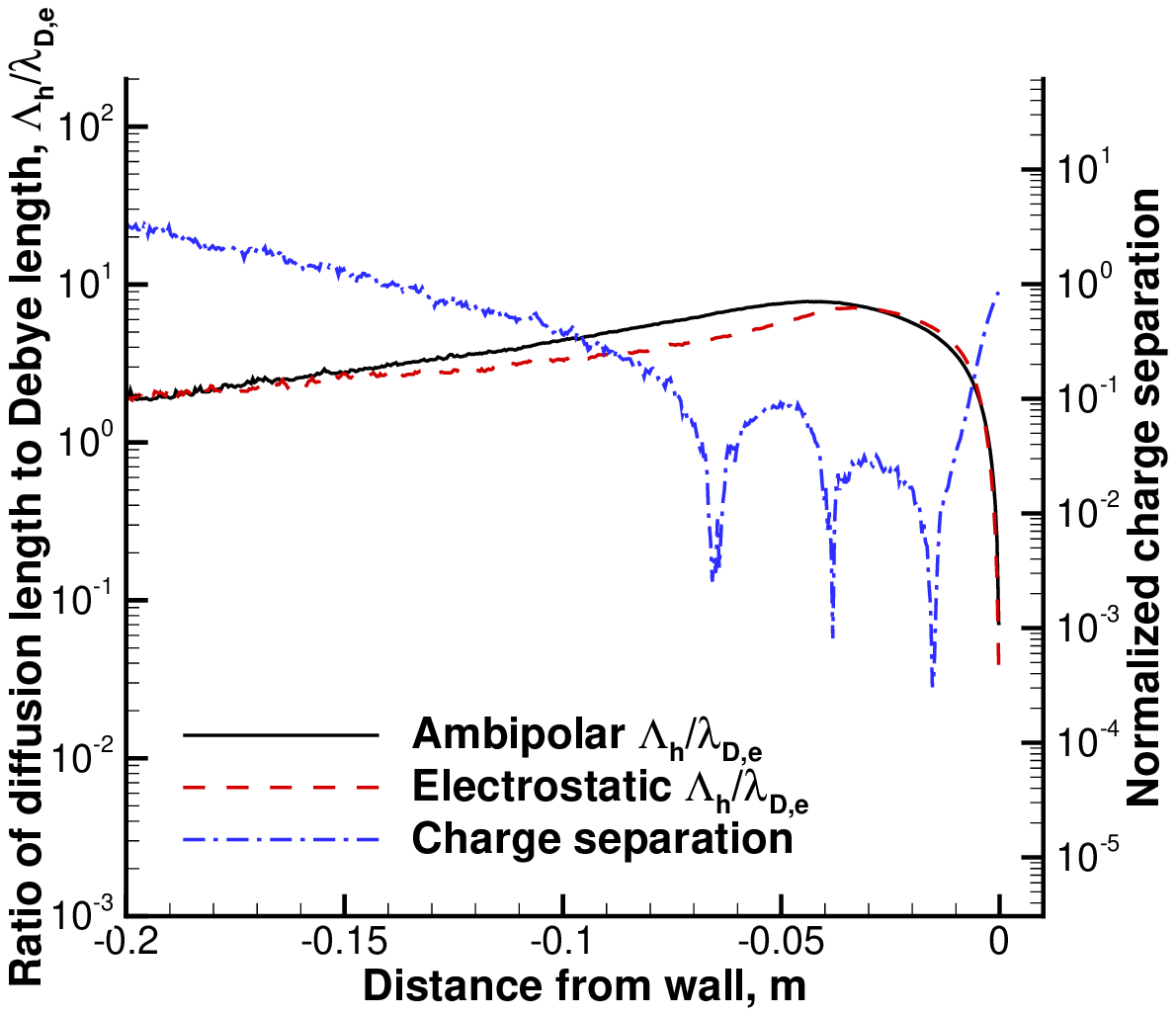}
			\caption{Transitional diffusion case ($F_r = 10^{8}$).}
			\label{fig:81km_Fr=1e8_diff_length}
		\end{subfigure}
		\begin{subfigure}{0.32\linewidth}
			\includegraphics[width=\linewidth]{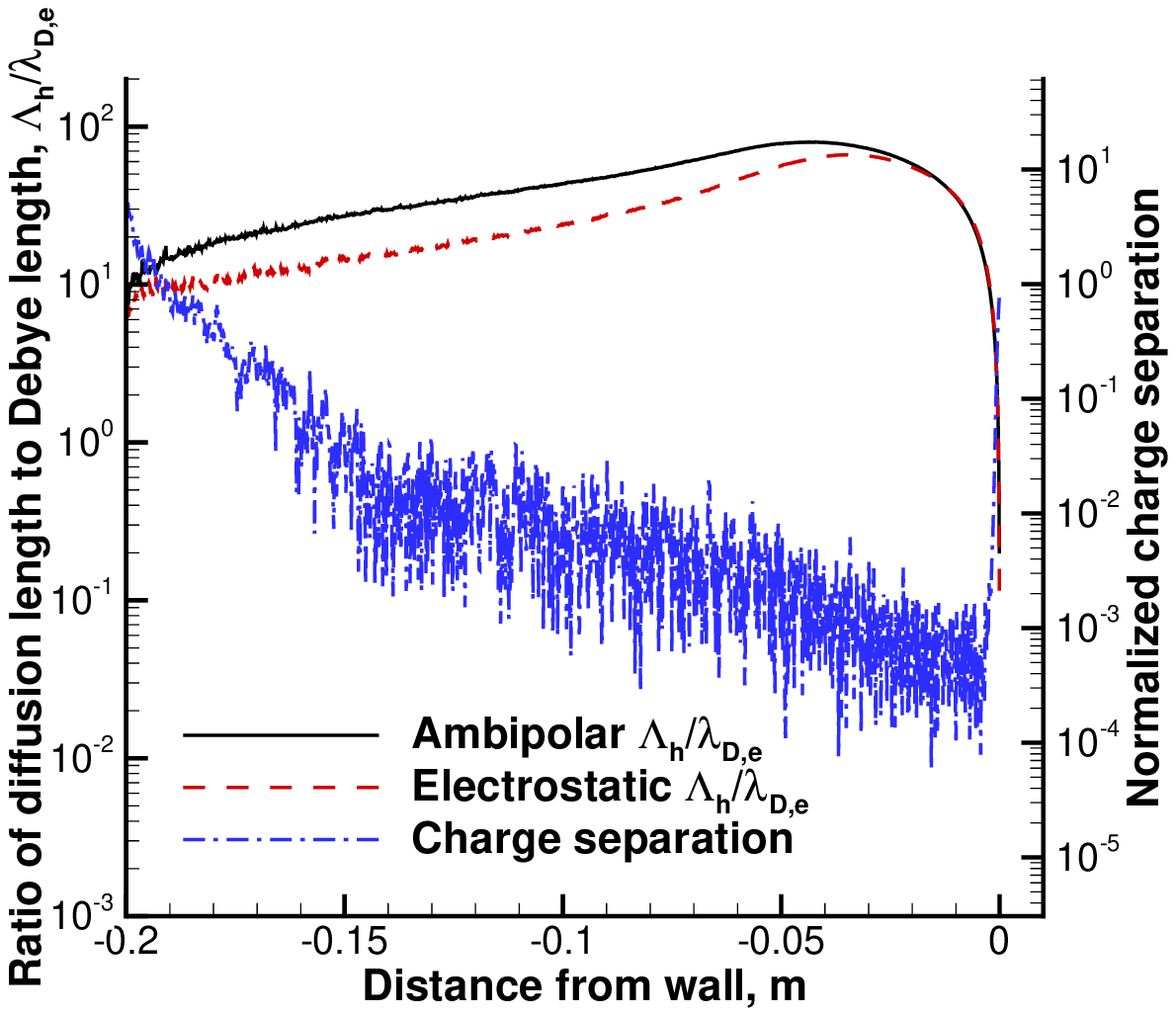}
			\caption{Ambipolar diffusion case ($F_r = 10^{7}$).}
			\label{fig:81km_Fr=1e7_diff_length}
		\end{subfigure}
		\caption{ Ratio of hypersonic diffusion length $\Lambda_h$ to electron Debye length $\lambda_{D,e}$ along a stagnation streamline with the ambipolar diffusion approximation and with full electrostatic modeling (Mach 39 flow, 81 km).}
		\label{fig:81km_diff_length}
	\end{figure*}

	Similar trends in electron temperature and heat fluxes are observed for the free diffusion case at Mach 39 and 81 km (Fig.~\ref{fig:81km_Fr=1e10}). The charge separation exceeds 10\% throughout the shock layer except from $-0.0517 < x <  -0.0409$ m and from $-0.0372 < x < -0.0190$ m, where it ranges between 1\% and 10\%. $\Lambda_h/\lambda_{D,e} < 1.097$ for the electrostatic solution, and $\Lambda_h/\lambda_{D,e} < 1.22$ for the ambipolar diffusion approximation solution (Fig.~\ref{fig:81km_Fr=1e10_diff_length}). Better agreement is achieved between the ion density profiles: the ambipolar diffusion approximation underpredicts the peak ion density by 5.09\% compared to 15.4\% in the 60 km flow case. The increased rarefaction of the freestream flow conditions results in less charge separation and more collective plasma behavior relative to the 60 km flow case due to fewer collisions with neutrals. For example, at the points of peak electron density, $\omega_{p,e}\tau_{e,n} = 1.01$ ($x = -0.00139$ m) and 2.60 ($x = -0.00986$ m) for the ambipolar and electrostatic solutions, respectively. This fulfills the third criterion for the definition of a plasma, but does not alone prove the partially ionized gas is a plasma nor exclude the gas from the free diffusion regime.
	
		\begin{table*}[tb]
			\centering
			\caption{Stagnation point heat flux by species for simulations of the free plasma diffusion regime in hypersonic flows. Each heat flux value is multiplied by $F_r$. The change is calculated relative to the ambipolar diffusion approximation value.}
			\label{tab:ch5_free_diff_heat_flux}
			\begin{tabular}{lccccccccc}
				\hline
				\hline
				& \multicolumn{3}{c}{Mach 35 at 60 km, $F_r = 10^{10}$} & \multicolumn{3}{c}{Mach 39 at 81 km, $F_r = 10^{10}$} \\
				Heat flux, W/cm$^2$ & Ambipolar & Electrostatic & Change & Ambipolar & Electrostatic & Change \\
				\hline
				Ar, convective &6,160 & 5,520 &$-$10.4\% & 401 & 373 & $-$6.98\%\\
				Ar$^+$, convective & 566 & 519 & $-$8.30\%&25.4 & 24.0 & $-$5.51\% \\
				Ar$^+$, chemical & 1,730 &  1,770& $+$2.31\%&88.5 & 84.4 & $-$4.63\% \\
				e$^-$, convective & 13.5 & 26.3 & $+$94.8\% &0.646 & 0.944 & $+$46.1\% \\
				Total heat flux, plasma & 2,310  &2,320 & $+$0.433\%& 115 & 109 & $-$5.22\%\\
				Percentage of total, plasma & 27.3\%& 29.5\%& $+$2.20\%& 22.3\%  & 22.6\% & $+$0.300\%\\
				Total heat flux, all species & 8,470 & 7,840 & $-$7.44\%& 516 &   482 & $-$6.59\% \\
				\hline
				\hline
			\end{tabular}
		\end{table*}
	
	\subsection{Transitional Diffusion}
	\label{sec:transitional_diff}
	
	\begin{figure}[tb]
		\centering 
		\begin{subfigure}{0.49\linewidth}
			\includegraphics[width=\linewidth]{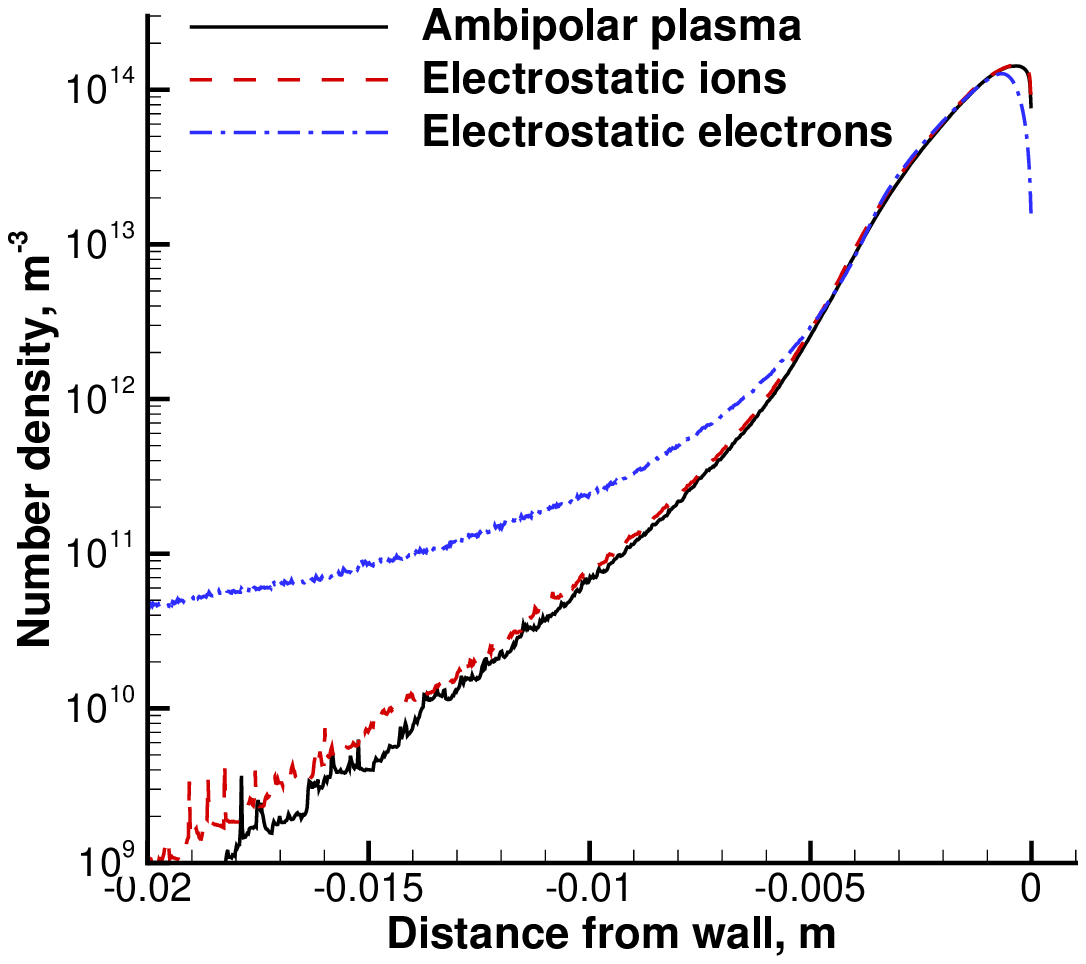}
			\caption{Charged species densities.}
			\label{fig:60km_Fr=1e8_dens}
		\end{subfigure}
		\begin{subfigure}{0.49\linewidth}
			\includegraphics[width=\linewidth]{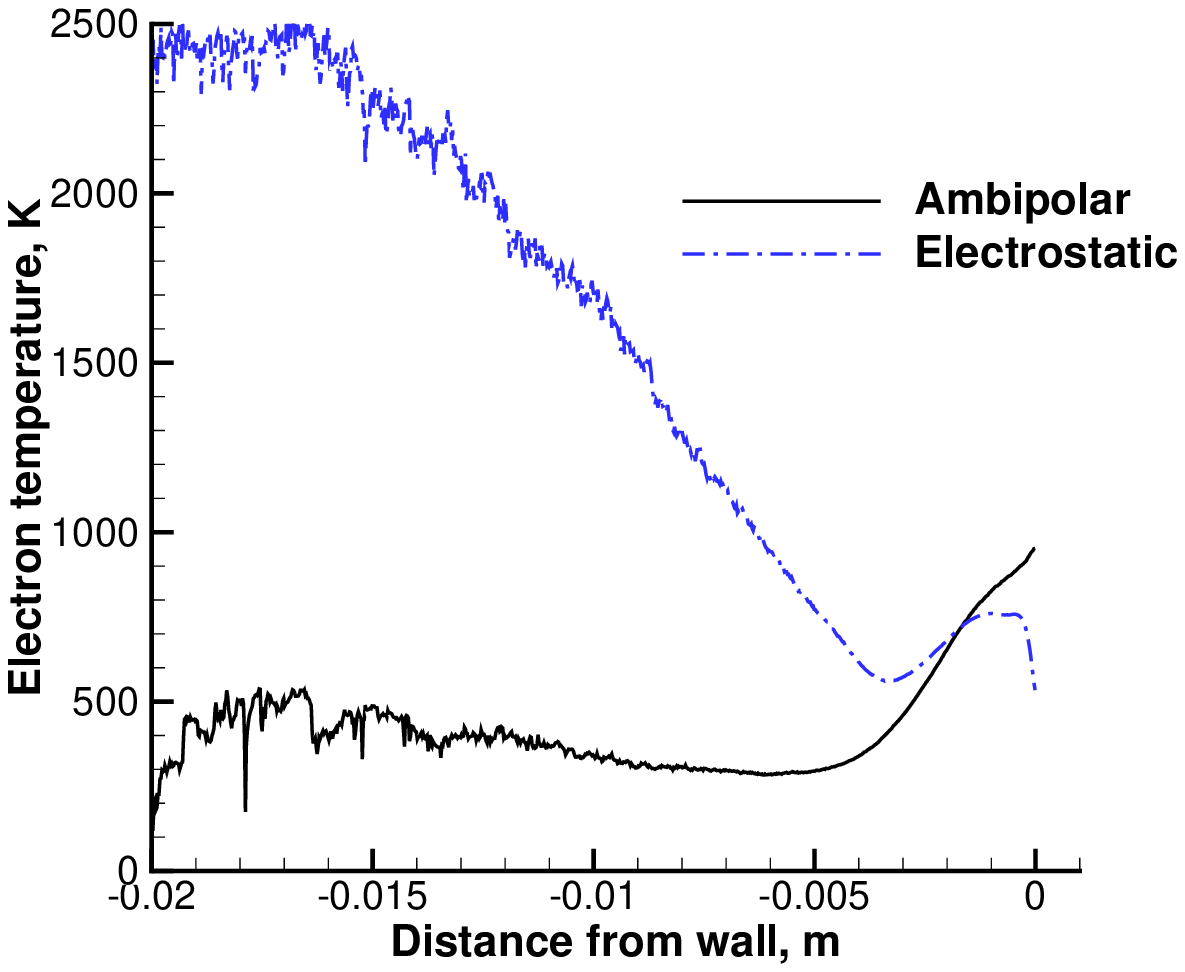}
			\caption{Electron temperature.}
			\label{fig:60km_Fr=1e8_elec_temp}
		\end{subfigure}
		\caption{Stagnation streamline flowfield under transitional plasma diffusion conditions with the ambipolar diffusion approximation and with full electrostatic modeling (Mach 35 flow, 60 km,  $F_r = 10^{8}$).}
		\label{fig:60km_Fr=1e8}
	\end{figure} 
	
	This work defines the transitional plasma diffusion regime for hypersonic flows as having normalized charge separation between 1\% and 10\% for a majority of the post-shock region, as the partially ionized gas is in an intermediary state between free and ambipolar diffusion. The Mach 35 case at 60 km uses $F_r = 10^8$, increasing the freestream density by a factor of 100 relative to the free diffusion case, while the Mach 39 case at 81 km uses $F_r = 10^9$, increasing the freestream density by a factor of 10. For the Mach 35 flow at 60 km, excluding the plasma sheath, the normalized charge separation ranges between 0.326\% and 9.26\%, with $\omega_{p,e} \tau_{e,n} = 6.22$ at the point of peak electrostatic electron density ($x = -0.000668$ m) and $\omega_{p,e} \tau_{e,n} = 2.80$ at the point of peak ambipolar plasma density ($x = -0.000351$ m) (Fig.~\ref{fig:60km_Fr=1e8_diff_length}). The profile of $\Lambda_h/\lambda_{D,e}$ from the electrostatic solution ranges between 3.30 and 9.75 throughout the shock layer, and decreases to 1.26 as one moves upstream: this aligns with the limits for transitional diffusion \cite{petrusky_evaluation_2026} where $1 < \Lambda/\lambda_{D,e} < 100$. For the ambipolar diffusion approximation solution, $\Lambda_h/\lambda_{D,e}$ ranges between 3.51 and 10.20 within the shock layer and decreases to 0.322 as one moves upstream. The electrostatic electron temperature is lower and flatter within the shock layer, except near the wall where it sharply decreases. Not shown, the value of $\vec{E} \cdot \vec{J}_e$ decreases from 0.934 W/m$^3$ at $x = -0.0283$ m to -7.57 W/m$^3$ at the wall. This is evidence of the electron cooling effect, where the increased plasma density in the transitional diffusion regime leads to stronger electric fields in the plasma sheath, increasing the magnitude of $\vec{E} \cdot \vec{J}_e$. Upstream of the shock, $\vec{E} \cdot \vec{J}_e$ is less than 10$^{-3}$ W/m$^3$, and thus is not responsible for the increased electron temperature. Similar to the free diffusion flow cases, the neutral and ion convective heat fluxes decrease with electrostatic modeling, and the electron convective heat flux increases, leading to net decreases in total stagnation point heat flux (Table \ref{tab:ch5_trans_diff_heat_flux}). Of note, the increase in electron heat flux ($+27.2\%$) is smaller than the free diffusion case ($+94.8\%$): this results from electron cooling lowering the average kinetic energy of electrons and fewer electrons reaching the surface due to the increased strength of the plasma sheath. 
	
	\begin{figure*}[tbh]
		\centering 
		\begin{subfigure}{0.49\linewidth}
			\includegraphics[width=\linewidth]{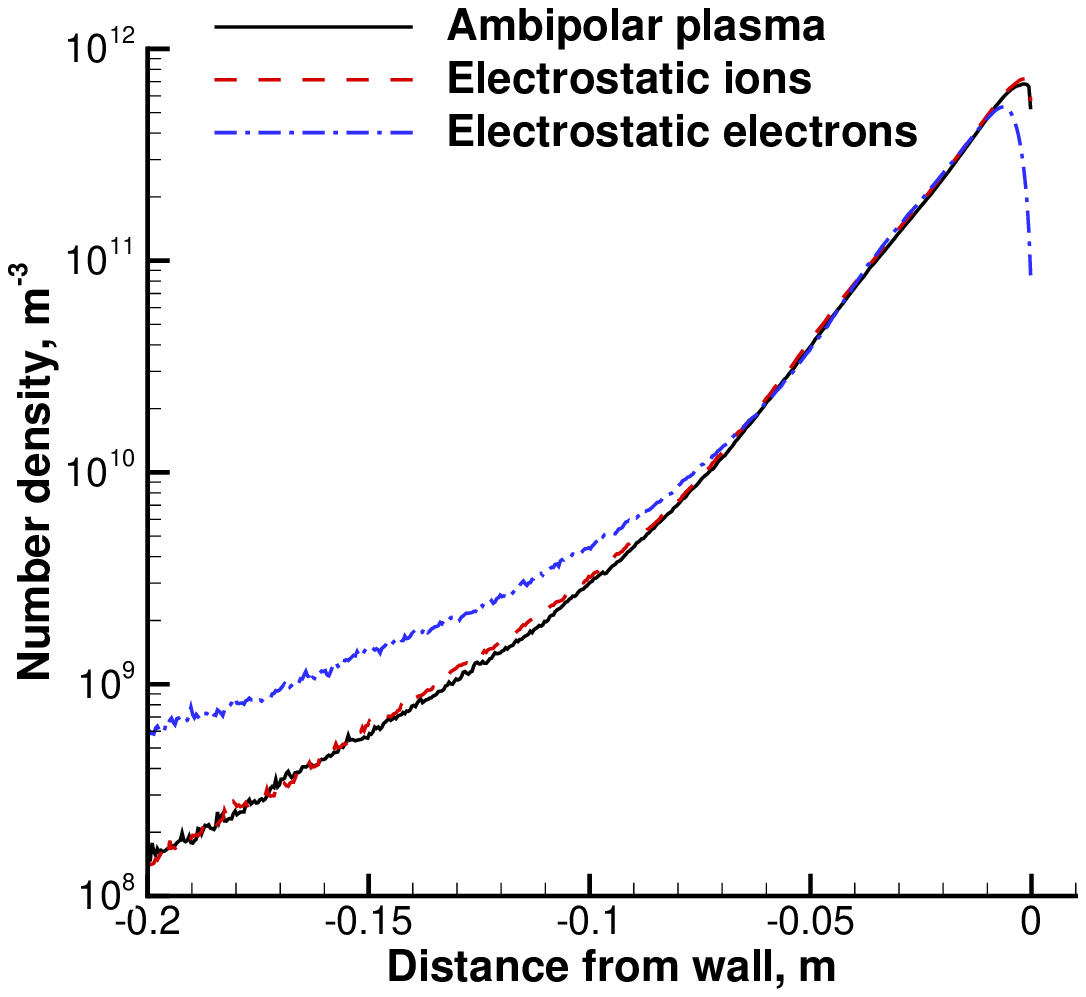}
			\caption{Charged species densities.}
			\label{fig:81km_Fr=1e8_dens}
		\end{subfigure}
		\begin{subfigure}{0.49\linewidth}
			\includegraphics[width=\linewidth]{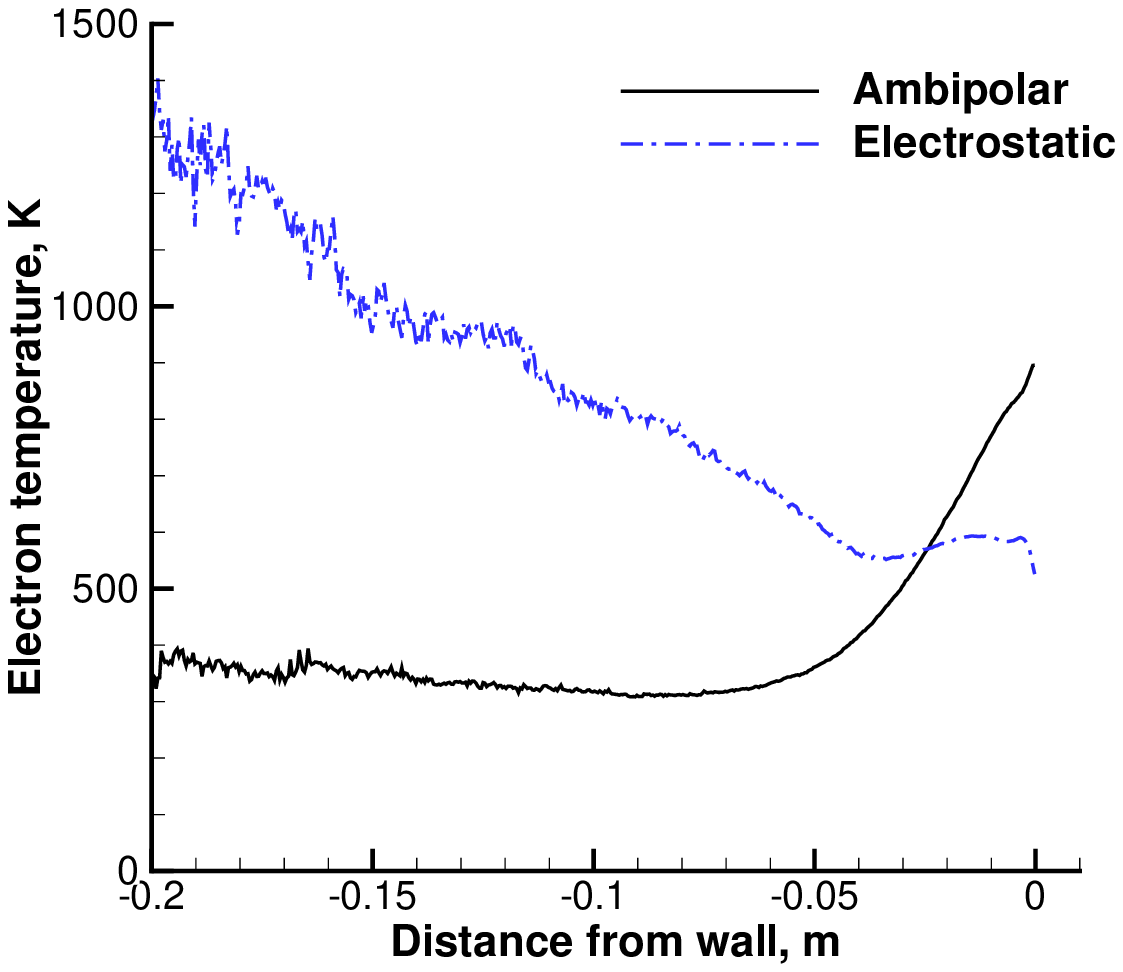}
			\caption{Electron temperature.}
			\label{fig:81km_Fr=1e8_elec_temp}
		\end{subfigure}
		\caption{Stagnation streamline flowfield under transitional plasma diffusion conditions with the ambipolar diffusion approximation and with full electrostatic modeling (Mach 39 flow, 81 km,  $F_r = 10^{9}$).}
		\label{fig:81km_Fr=1e8}
	\end{figure*} 
	
	Figure~\ref{fig:81km_Fr=1e8} presents transitional diffusion flowfield results with the ambipolar diffusion approximation and with full electrostatic modeling for the Mach 39 flow at 81 km. Throughout the post-shock region (excluding the plasma sheath), the charge separation ranges between 0.0373\% and 2.60\%, $(\Lambda_h/\lambda_{D,e})_{max} = 7.12$ at $x = -0.0342$ m for the electrostatic solution and $(\Lambda_h/\lambda_{D,e})_{max} = 7.78$ at $x = -0.0412$ m for the ambipolar diffusion approximation (Fig.~\ref{fig:81km_Fr=1e8_diff_length}), and $\omega_{p,e} \tau_{e,n} = 8.11$ at the point of peak electrostatic electron density ($x = -0.00663$ m) and $\omega_{p,e} \tau_{e,n} = 3.23$ at the point of peak ambipolar plasma density ($x = -0.00199$ m). These properties are all characteristic of the transitional plasma diffusion regime. A sharp decrease in electron temperature occurs at the wall, despite the fact that $\vec{E} \cdot \vec{J}_e$ ranges from 1.99 $\times 10^{-4}$ W/m$^{3}$ at $x = -0.014$ m to $-2.37 \times 10^{-3}$ W/m$^{3}$ at the wall. Because the flow is more rarefied and fewer collisions with neutrals occur, a smaller magnitude of $\vec{E} \cdot \vec{J}_e$ can lead to electron cooling. Overall, the differences in stagnation point heat flux are negligible with the exception of the electron convective heat flux.
	
	\begin{table*}
		\centering
		\caption{Stagnation point heat flux by species for simulations of the transitional plasma diffusion regime in hypersonic flows. Each heat flux value is multiplied by $F_r$. The change is calculated relative to the ambipolar diffusion approximation value.}
		\label{tab:ch5_trans_diff_heat_flux}
		\begin{tabular}{lcccccc}
			\hline
			\hline
			& \multicolumn{3}{c}{60 km, $F_r = 10^8$}
			& \multicolumn{3}{c}{81 km, $F_r = 10^9$}\\
			Heat flux, W/cm$^2$ & Ambipolar & Electrostatic & Change
			& Ambipolar & Electrostatic & Change \\
			\hline
			Ar, convective & 6,170 & 5,690 & $-$7.78\%
			& 394 & 381 & $-$3.30\%\\
			Ar$^+$, convective & 574 & 482 & $-$16.0\%
			& 25.4 & 26.3 & $+$3.54\%\\
			Ar$^+$, chemical & 1,750 & 1,640 & $-$6.29\%
			& 86.3 & 89.2 & $+$3.36\%\\
			e$^-$, convective & 13.6 & 17.3 & $+$27.2\%
			& 0.634 & 0.909 & $+$43.4\%\\
			Total heat flux, plasma & 2,340 & 2,140 & $-$8.55\%
			& 112 & 116 & $+$3.57\%\\
			Percentage of total, plasma & 27.5\% & 27.3\% & $-$0.200\%
			& 22.2\% & 23.3\% & $+$1.10\%\\
			Total heat flux, all species & 8,510 & 7,830 & $-$7.99\%
			& 506 & 497 & $-$1.78\%\\
			\hline
			\hline
		\end{tabular}
	\end{table*}

	\subsection{Ambipolar Diffusion}
	\label{sec:ambi_diff}
	
	\begin{figure}[htb]
		\centering 
		\begin{subfigure}{0.49\linewidth}
			\includegraphics[width=\linewidth]{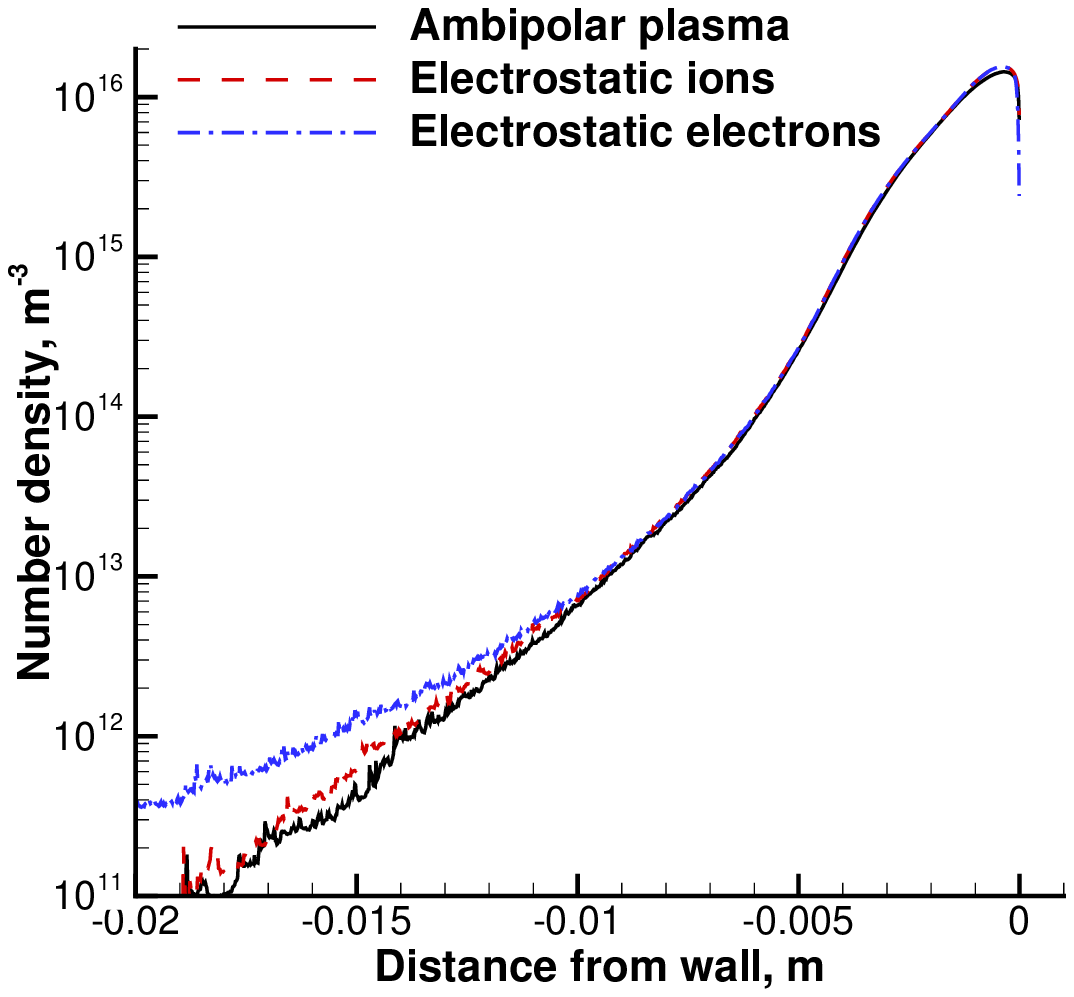}
			\caption{Charged species densities.}
			\label{fig:60km_Fr=1e6_dens}
		\end{subfigure}
		\begin{subfigure}{0.49\linewidth}
			\includegraphics[width=\linewidth]{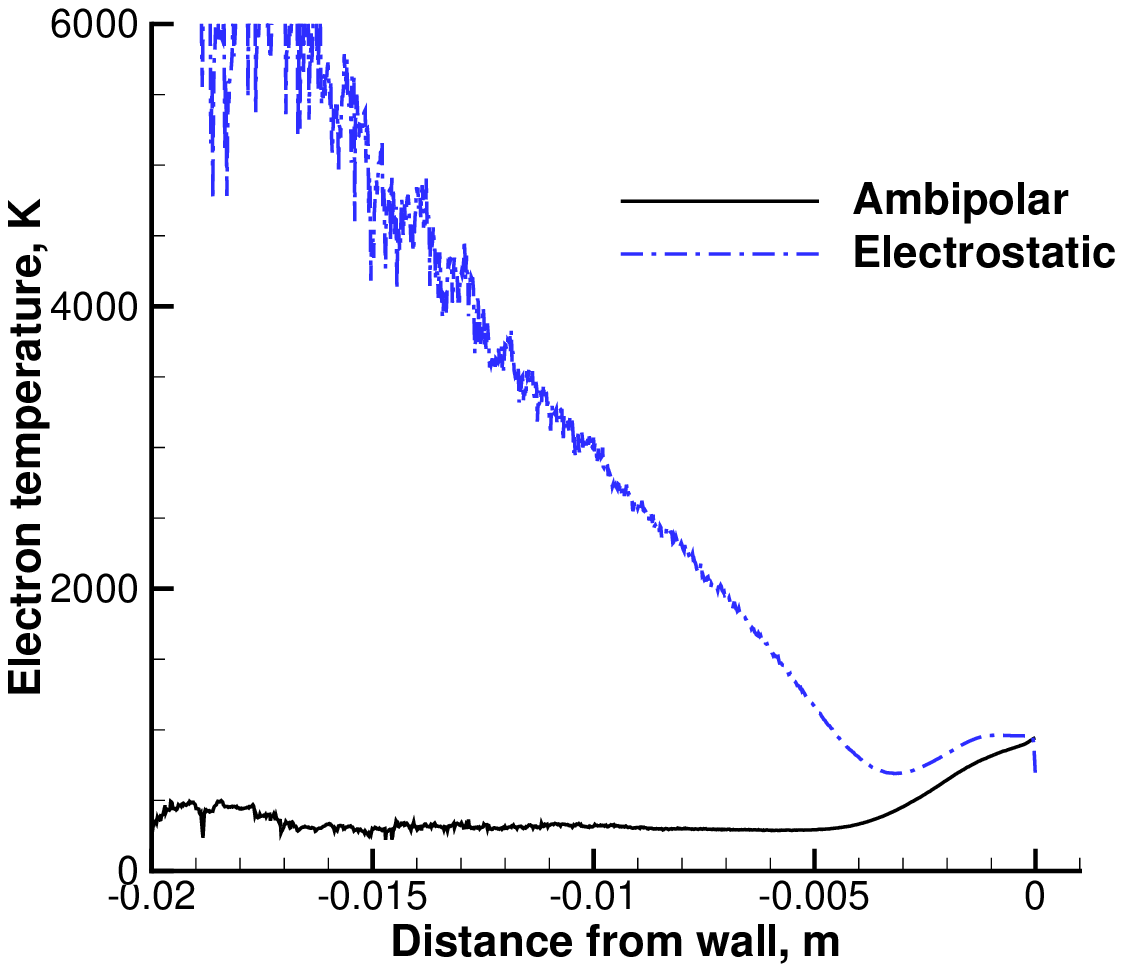}
			\caption{Electron temperature.}
			\label{fig:60km_Fr=1e6_elec_temp}
		\end{subfigure}
		\caption{Stagnation streamline flowfield under ambipolar plasma diffusion conditions with the ambipolar diffusion approximation and with full electrostatic modeling (Mach 35 flow, 60 km,  $F_r = 10^{6}$).}
		\label{fig:60km_Fr=1e6}
	\end{figure} 
		
	\begin{figure}[tb]
		\centering 
		\includegraphics[width=0.49\linewidth]{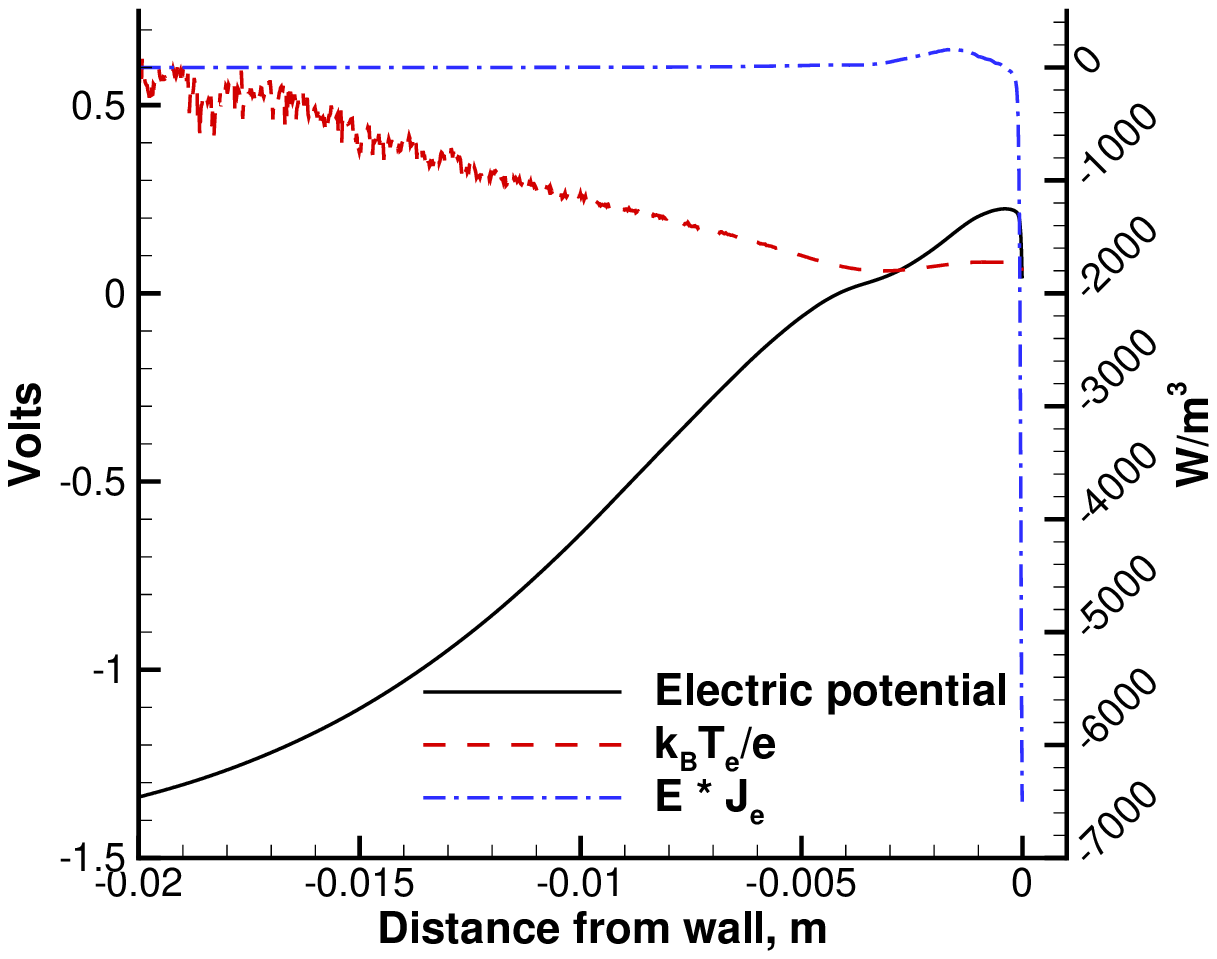}
		\caption{(Left axis) Electric potential and electron thermal energy per unit charge; (Right axis) Dot product of the electric field and the electron current density (Mach 35 flow, 60 km,  $F_r = 10^{6}$).}
		\label{fig:60km_Fr=1e6_other}
	\end{figure} 
	
	This work defines the ambipolar diffusion regime for hypersonic flows as having a majority of the post-shock region undergo ambipolar diffusion with charge separation below 1\%. Implicitly, the post-shock partially ionized gas must be classified as a plasma. Figure~\ref{fig:60km_Fr=1e6} presents flowfield results with the ambipolar diffusion approximation and with full electrostatic modeling for the Mach 35 flow at 60 km when $F_r = 10^6$. The post-shock plasma number density is on the order of $10^{16}$ m$^{-3}$, a realistic plasma density for certain hypersonic flow conditions \cite{monroe_electron_2025}. As Fig. \ref{fig:60km_Fr=1e6_other} shows, the average electron thermal energy per unit charge is roughly half that of the electric potential within the shock layer; thus few electrons are able to diffuse out of the shock layer without ions to accompany them. Upstream of the shock, the normalized charge separation exceeds 1\% at $x = -0.00927$ m, where $\Lambda_h/\lambda_{D,e} = 24.3$ for the ambipolar solution and $\Lambda_h/\lambda_{D,e} = 8.7$ for the electrostatic solution. The $\Lambda_h/\lambda_{D,e}$ profiles reach their maxima at the shock standoff distance, with $\Lambda_h/\lambda_{D,e} = $104 and 86.4 for the ambipolar and electrostatic solutions, respectively. Under ambipolar diffusion conditions, the ambipolar diffusion coefficient can be approximated as\cite{ChenTB1984}
		
	\begin{equation}
		\label{eq:diff_coeff}
		D_a \approx D_i (1 + T_e/T_i)
	\end{equation}
		
	\noindent where $D_i$ is the diffusion coefficient of ions. The ion temperature is on the order of 100,000 K throughout the shock layer, similar to the neutrals. Therefore, $D_a \approx D_i$ for $T_e \ll T_i$. Since the neutral density distributions are relatively unchanged between the ambipolar diffusion approximation and electrostatic solutions, the models share the same $D_i \approx D_a$ and the ambipolar plasma density agrees strongly with the electrostatic ion and electron densities. The approximation is still unable to capture the distinct electrostatic electron temperature profile; in particular, the electrostatic electron temperature profile is greater than the ambipolar diffusion approximation's throughout a majority of the shock layer. From $-0.0159 < x < -0.000410$ m (Fig.~\ref{fig:60km_Fr=1e6_other}), $\vec{E} \cdot \vec{J}_e > 0$, with a maximum positive value of 157 W/m$^3$ at $x = -0.00146$ m. This leads to electron heating upstream of the shock. $\omega_{p,e}\tau_{e,n} = 36.4$ at the point of peak electrostatic electron density ($x = -0.000411$ m), and $\omega_{p,e}\tau_{e,n} = 37.3$ at the point of peak ambipolar plasma density ($x = -0.000370$ m). In terms of species-specific stagnation point heat flux quantities (Table \ref{tab:ch5_ambi_diff_heat_flux}), the neutral, ion, and electron convective heat fluxes demonstrate similar trends to the free and transitional diffusion flow cases. The decrease in total heat flux is the smallest among the 60 km flow cases with $-5.52\%$.

		\begin{figure}[htb]
	\centering 
	\begin{subfigure}{0.49\linewidth}
		\includegraphics[width=\linewidth]{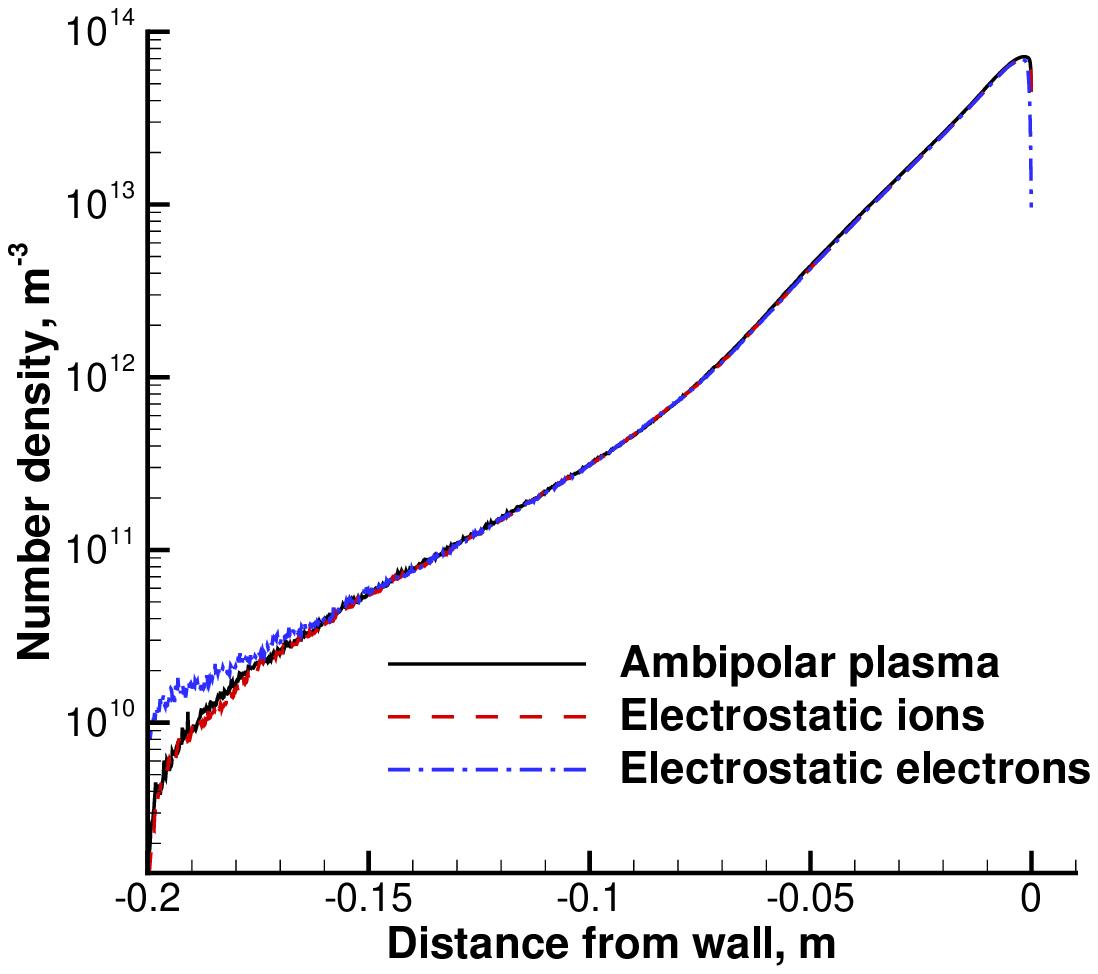}
		\caption{Charged species densities.}
		\label{fig:81km_Fr=1e7_dens}
	\end{subfigure}
	\hfill
	\begin{subfigure}{0.49\linewidth}
		\includegraphics[width=\linewidth]{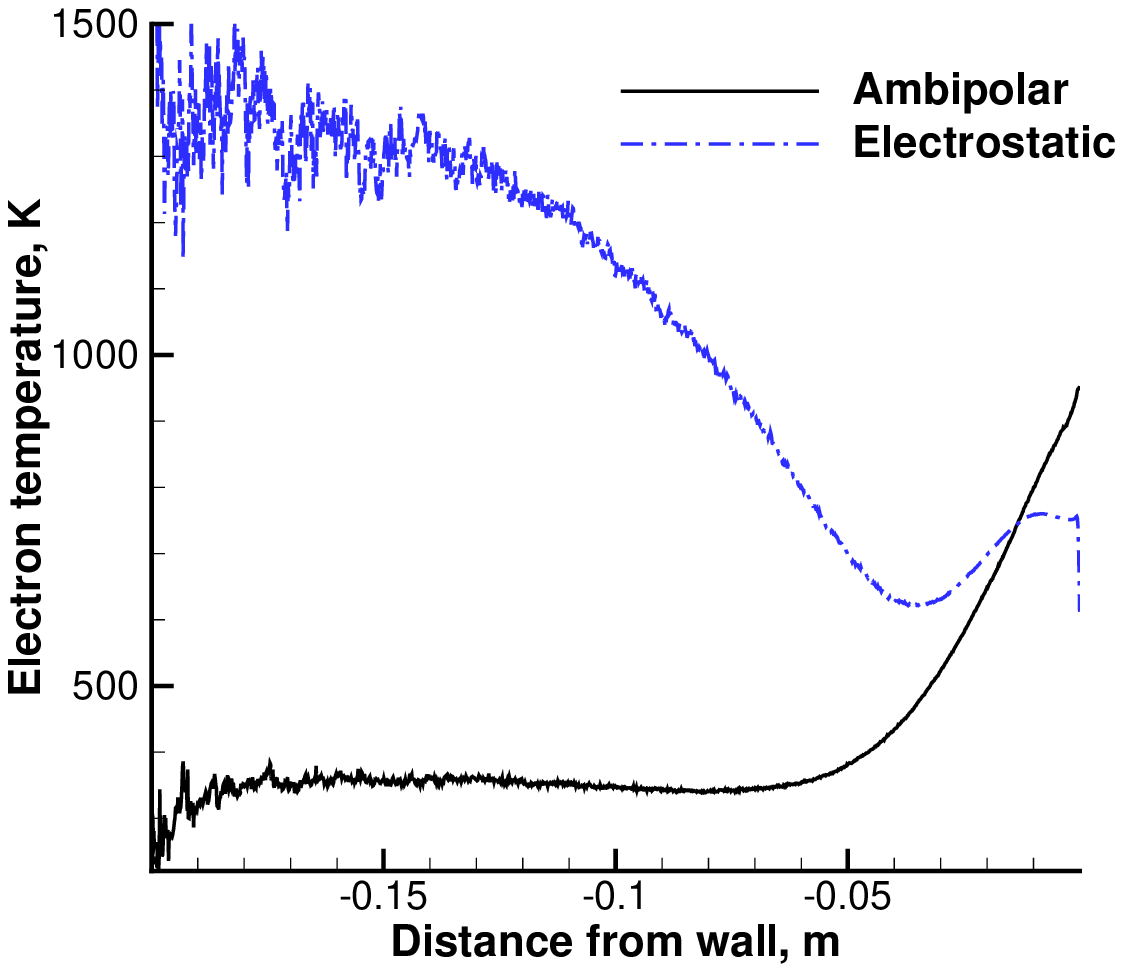}
		\caption{Electron temperature.}
		\label{fig:81km_Fr=1e7_elec_temp}
	\end{subfigure}
	\caption{Stagnation streamline flowfield under ambipolar plasma diffusion conditions with the ambipolar diffusion approximation and with full electrostatic modeling (Mach 39 flow, 81 km,  $F_r = 10^{7}$).}
	\label{fig:81km_Fr=1e7}
\end{figure} 

	Figure~\ref{fig:81km_Fr=1e7} presents flowfield results with the ambipolar diffusion approximation and with full electrostatic modeling for the Mach 39 flow at 81 km when $F_r = 10^7$. Outside of the plasma sheath, the charge separation is below 1\% up until $x = -0.145$ m, where $\Lambda_h/\lambda_{D,e} = 15.0$ for the electrostatic solution and $\Lambda_h/\lambda_{D,e} = 28.4$ for the ambipolar diffusion approximation solution. $\omega_{p,e}\tau_{e,n} = 44.1$ at the point of peak electrostatic electron density and $\omega_{p,e}\tau_{e,n} = 43.8$ at the point of peak ambipolar plasma density; both density maxima are located at $x = -0.00237$ m. $(\Lambda_h/\lambda_{D,e})_{max} = 66.1$ at $x = -0.0328$ m for the electrostatic solution and $(\Lambda_h/\lambda_{D,e})_{max} = 79.6$ at $x = -0.0424$ m for the ambipolar diffusion approximation (Fig.~\ref{fig:81km_Fr=1e7_diff_length}). Thus, the flowfield meets the criteria for ambipolar diffusion. Electron cooling and heating are present, although the electrostatic electron temperature profile does not exceed the ambipolar diffusion approximation's in the post-shock region. The heat fluxes have negligible changes between the ambipolar and electrostatic solutions, with the exception of the electron convective heat flux which increases by $27.2\%$ with electrostatic modeling. 
	
	\begin{table*}
		\centering
		\caption{Stagnation point heat flux by species for simulations of the ambipolar plasma diffusion regime in hypersonic flows. Each heat flux value is multiplied by $F_r$. The change is calculated relative to the ambipolar diffusion approximation value.}
		\label{tab:ch5_ambi_diff_heat_flux}
		\begin{tabular}{lcccccc}
			\hline
			\hline
			& \multicolumn{3}{c}{60 km, $F_r = 10^6$} & \multicolumn{3}{c}{81 km, $F_r = 10^7$}\\
			Heat flux, W/cm$^2$ & Ambipolar & Electrostatic & Change & Ambipolar & Electrostatic & Change \\
			\hline
			Ar, convective & 6,280 & 5,850 & $-$6.85\% & 377 & 378 &$+$0.26\% \\
			Ar$^+$, convective & 600 & 539 & $-$10.2\% & 24.3 & 23.6 & $-$2.88\%\\
			Ar$^+$, chemical & 1,800 & 1,800 &$+$0.00\% & 89.9 & 89.5 & $-$0.445\% \\
			e$^-$, convective & 13.9 & 21.6 & $+$55.4\% & 0.702& 0.893 & $+$27.2\%\\
			Total heat flux, plasma &2,410 & 2,360& $-$2.07\% & 115 & 114 & $-$0.870\% \\
			Percentage of total, plasma & 27.7\%& 28.7\%& $+$1.00\% & 23.4\% & 23.2\% & $-$0.200\% \\
			Total heat flux, all species & 8,690& 8,210 & $-$5.52\% & 492 & 492 & $+$0.00\% \\
			\hline
			\hline
		\end{tabular}
	\end{table*}

	\subsection{Electron Mass Sensitivity Analysis}
	\label{sec:mass_sens}
	
	Artificially reducing the electron mass will lead to differences in ionization rates, electron collision rates, flowfield distributions, and diffusion behavior across the same flow case. Therefore, it is necessary to assess how the trends observed in this study may change with different values of $m_e$. The transitional diffusion flow case at Mach 39 and 81 km is run with $m_i/m_e = 500$ and $m_i/m_e =$ 1,500, and then compared to the results in Section \ref{sec:transitional_diff} with $m_i/m_e =$ 1,000.
	
\begin{figure*}[tbh]
		\centering 
		\begin{subfigure}{0.49\linewidth}
			\includegraphics[width=\linewidth]{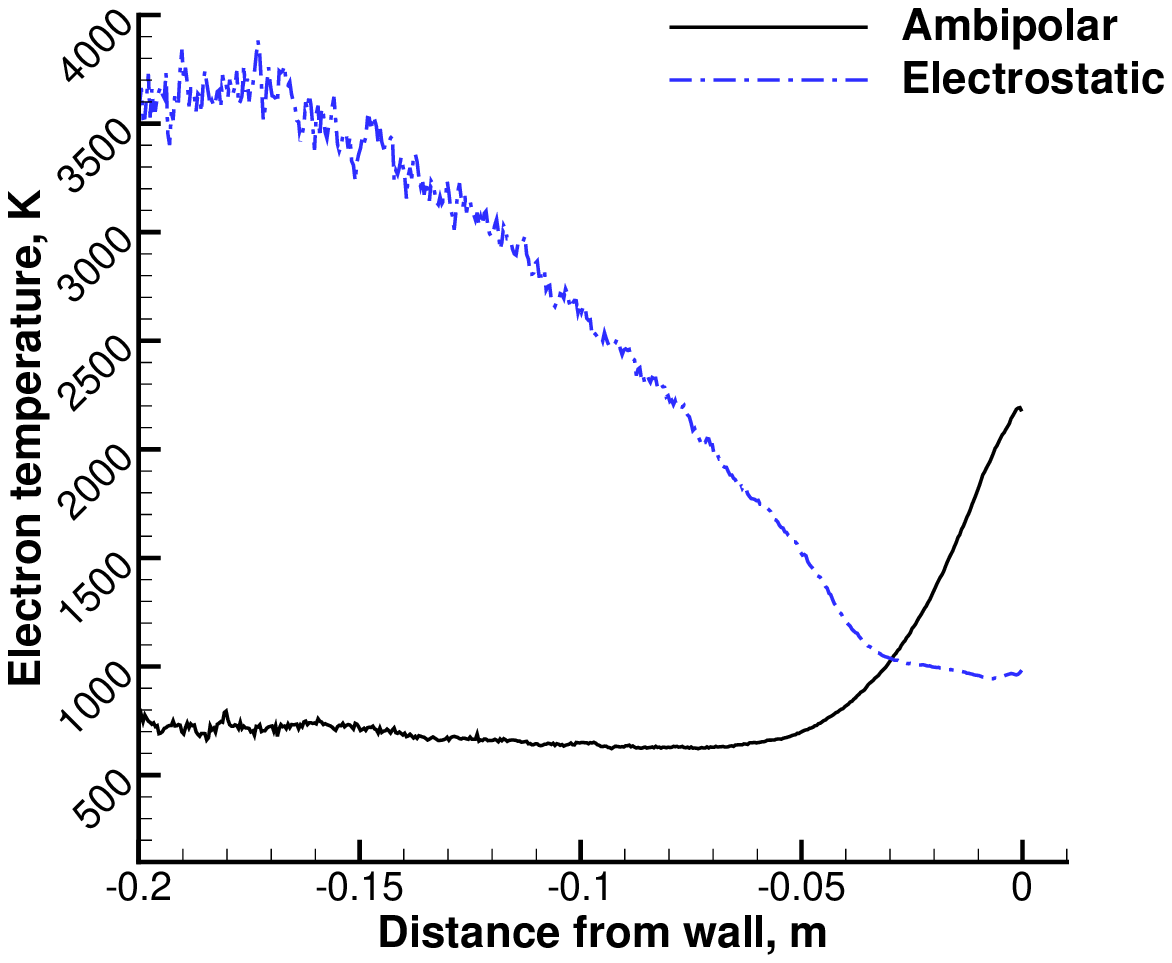}
			\caption{$m_i/m_e = 500$.}
			\label{fig:81km_Fr=1e10_elec_temp_me/mi=100}
		\end{subfigure}
		\begin{subfigure}{0.49\linewidth}
			\includegraphics[width=\linewidth]{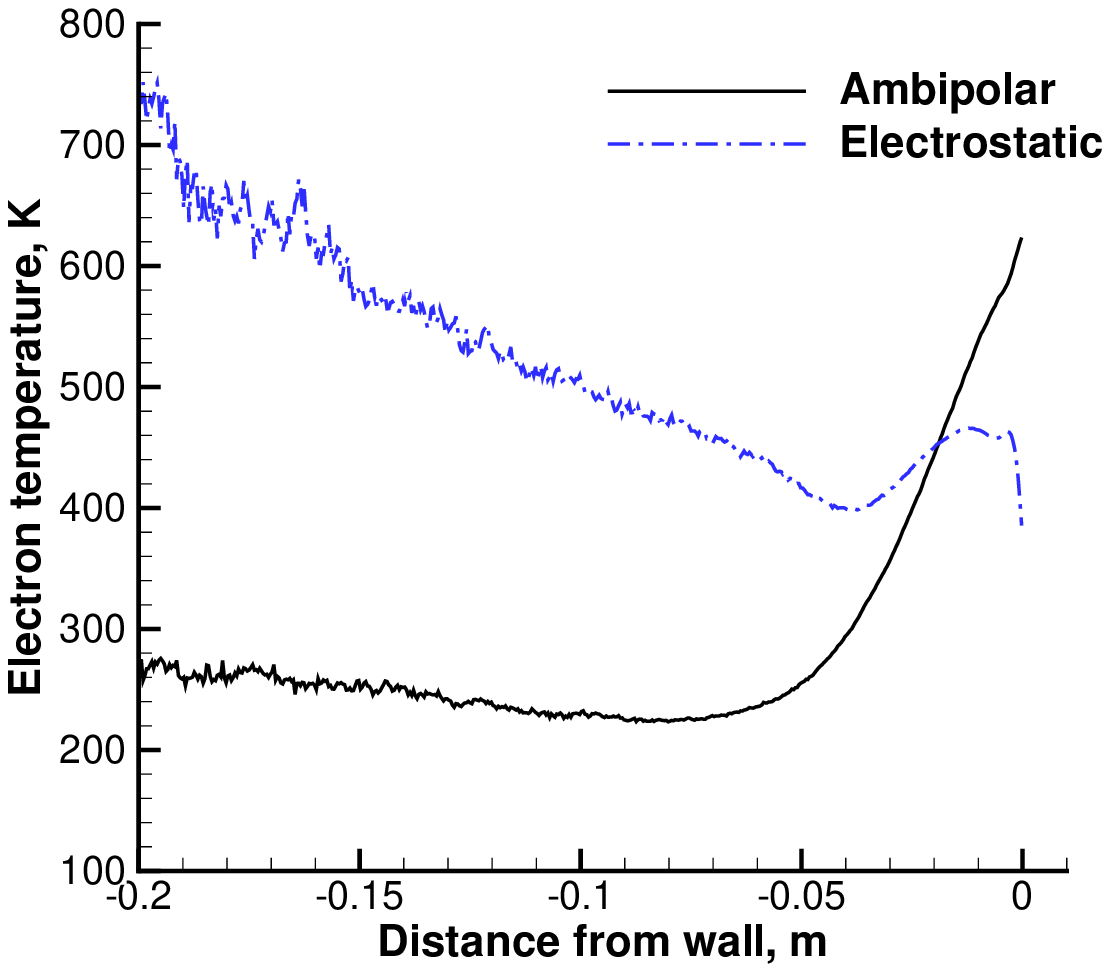}
			\caption{$m_i/m_e = 1500$.}
			\label{fig:81km_Fr=1e10_dens_me/mi=100}
		\end{subfigure}
		\caption{Electron temperature profile under transitional plasma diffusion conditions with the ambipolar diffusion approximation and with full electrostatic modeling for different ratios of $m_i/m_e$ (Mach 39 flow, 81 km,  $F_r = 10^{9}$).}
		\label{fig:mass_elec_temps}
	\end{figure*}

	For the $m_i/m_e = 500$ flow case, $\Lambda_h/\lambda_{D,e}$ reaches a maximum value of 5.07 for the electrostatic solution and 5.85 for the ambipolar diffusion approximation solution. The charge separation ranges between 0.251 and 3.72\% in the post-shock region. When a higher electron mass is used, electrons receive a larger fraction of energy per neutral-electron collision, leading to higher electron temperatures throughout the domain. As seen in Fig.~\ref{fig:81km_Fr=1e10_elec_temp_me/mi=100}, the electron temperature at the shock standoff distance ($x = -0.0243$ m) increases from 592 K in the $m_i/m_e = $1,000 case to 984 K. The steep drop in electron temperature associated with electron cooling does not manifest because $Q_{e,n}$ increases with a larger electron mass. From Eq.~(\ref{eq:diff_coeff}), an increased $T_e$ leads to enhanced transport of the electrostatic ions via the ambipolar diffusion coefficient. The ambipolar diffusion approximation overpredicts the peak ion density by 5.56\%, and the location of the density maximum shifts upstream by 0.684 mm with electrostatic modeling, indicating that more ions are diffusing out of the shock layer. For comparison, the ambipolar diffusion approximation \textit{underpredicts} the peak ion density by 5.65\% for the $m_i/m_e = 1,000$ flow case, and the location of the maximum shifts by 0.352 mm closer to the wall. Most of the stagnation point heat fluxes changes from the ambipolar diffusion approximation solution are negligible, except for the electron convective heat flux which increases by a similar amount to the $m_i/m_e =$ 1,000 flow case with electrostatic modeling.
	
	For the $m_i/m_e =$ 1,500 flow case, the charge separation ranges between 0.148\% and 6.96\% in the post-shock region, with a maximum value of $\Lambda_h/\lambda_{D,e} = $8.27 for the electrostatic solution and 9.12 for the ambipolar diffusion approximation solution. Lower electron temperatures are present throughout the domain, with $T_e = 436$ K at the shock standoff distance. Electron cooling is also present at the sheath. The ambipolar diffusion approximation underpredicts the peak ion density by 4.58\% relative to the electrostatic solution, and to 3 significant figures, the locations of the maxima are the same. Since $T_e$ decreases with increasing $m_i/m_e$, agreement between the electrostatic ion and ambipolar plasma density profiles improves since $D_a$ approaches $D_i$. Relative to the ambipolar diffusion approximation solution, the electrostatic neutral convective, ion convective, and total heat fluxes decrease by over 5\%, whereas the changes for the $m_i/m_e =$ 1,000 case are negligible. Similarly, the electron convective heat flux increases by 36.6\% with electrostatic modeling, compared to $+43.4$\% with $m_i/m_e =$ 1,000.
	
		\begin{table*}
		\centering
		\caption{Stagnation point heat flux by species for simulations of the transitional plasma diffusion regime in hypersonic flows with different ratios of $m_i/m_e$. Each heat flux value is multiplied by $F_r = 10^9$. The change is calculated relative to the ambipolar diffusion approximation value.}
		\label{tab:mass_sens_heat_flux}
		\begin{tabular}{lcccccc}
			\hline
			\hline
			& \multicolumn{3}{c}{$m_i/m_e = 500$} & \multicolumn{3}{c}{$m_i/m_e = 1,500$}\\
			Heat flux, W/cm$^2$ & Ambipolar & Electrostatic & Change & Ambipolar & Electrostatic & Change \\
			\hline
			Ar, convective & 380 & 378 & $-$0.526\% & 404 & 382 &$-$5.44\% \\
			Ar$^+$, convective & 26.1 & 25.0 & $-$4.21\% & 26.6 & 25.2 & $-$5.26\%\\
			Ar$^+$, chemical & 90.0 & 86.3 &$-$4.11\% & 87.2 & 86.7 & $-$0.573\% \\
			e$^-$, convective & 1.62 & 2.06 & $+$27.2\% & 0.440& 0.601 & $+$36.6\%\\
			Total heat flux, plasma & 118 & 113& $-$4.23\% & 114 & 113 & $-$0.870\% \\
			Percentage of total, plasma & 23.7\%& 23.0\%& $-$0.70\% & 22.0\% & 22.8\% & $+$0.800\% \\
			Total heat flux, all species & 498 & 491 & $-$1.41\% & 518 & 495 & $-$4.44\% \\
			\hline
			\hline
		\end{tabular}
	\end{table*}
	
	\section{Discussion}
	\label{sec:summary}
	
	\subsection{Hypersonic Plasma Behavior}
	
	Across both hypersonic freestream conditions and all three charged species diffusion regimes, several trends in rarefied plasma behavior along a stagnation streamline are identified in this study:
	
	\begin{enumerate}
		\item A negative electric field forms along most of the streamline, reaching its peak value between the center of the shock and the shock standoff distance and increasing as one approaches either the inflow boundary or the wall. Within a sufficient distance of the wall, the electric field switches sign as it couples to the fields of the plasma sheath. 
		\item `Electron heating and cooling', first identified in Ref.~\onlinecite{parent_electron_2021}, where electrons lose kinetic energy if their bulk velocity points in the same direction as the ambipolar electric field, and gain kinetic energy if their bulk velocity points in the opposite direction according to Eq.~(\ref{eq:electron_energy_transport}). This results in decreased electron temperatures in the vicinity of the plasma sheath and increased electron temperatures upstream of the shock if energy exchange between neutrals and ions from collisions and electron-impact ionization is sufficiently small. This work is the first to demonstrate that electron heating and cooling can occur for flowfields undergoing transitional and ambipolar diffusion; and that the phenomena are not unique to continuum flows or continuum modeling (since this study models rarefied flows with a kinetic approach). 
		\item `Convective heat loss', where energy transfer via MEX collisions with independently moving charged species (particularly lightweight electrons) that diffuse out of the shock layer (either upstream or to the wall) results in a decreased neutral convective heat flux to the wall. For some flow conditions, this may result in a non-negligible \textit{decrease} in total stagnation point heat flux. The ion convective heat flux may also decrease under some flow conditions.
		\item `Increased electron convective heat flux', where across all flow cases, including the electron mass sensitivity study, the electron convective heat flux increases relative to the ambipolar diffusion approximation by over 20\%. By enabling independent motion of electrons (either through electrostatic modeling or the free diffusion approximation), larger gradients in electron density form at the wall, increasing diffusion. Because electrons have a lighter mass and typical hypersonic flows are not fully ionized, the increased electron convective heat flux does not have a significant impact on the total heat flux. However, increased transport of electrons to the wall is highly relevant for applications involving plasma sheaths, such as electron-transpiration cooling\cite{monroe_electron_2025}.
	\end{enumerate}
	
\begin{figure*}[tb]
		\centering 
			\begin{subfigure}{0.49\linewidth}
			\includegraphics[width=\linewidth]{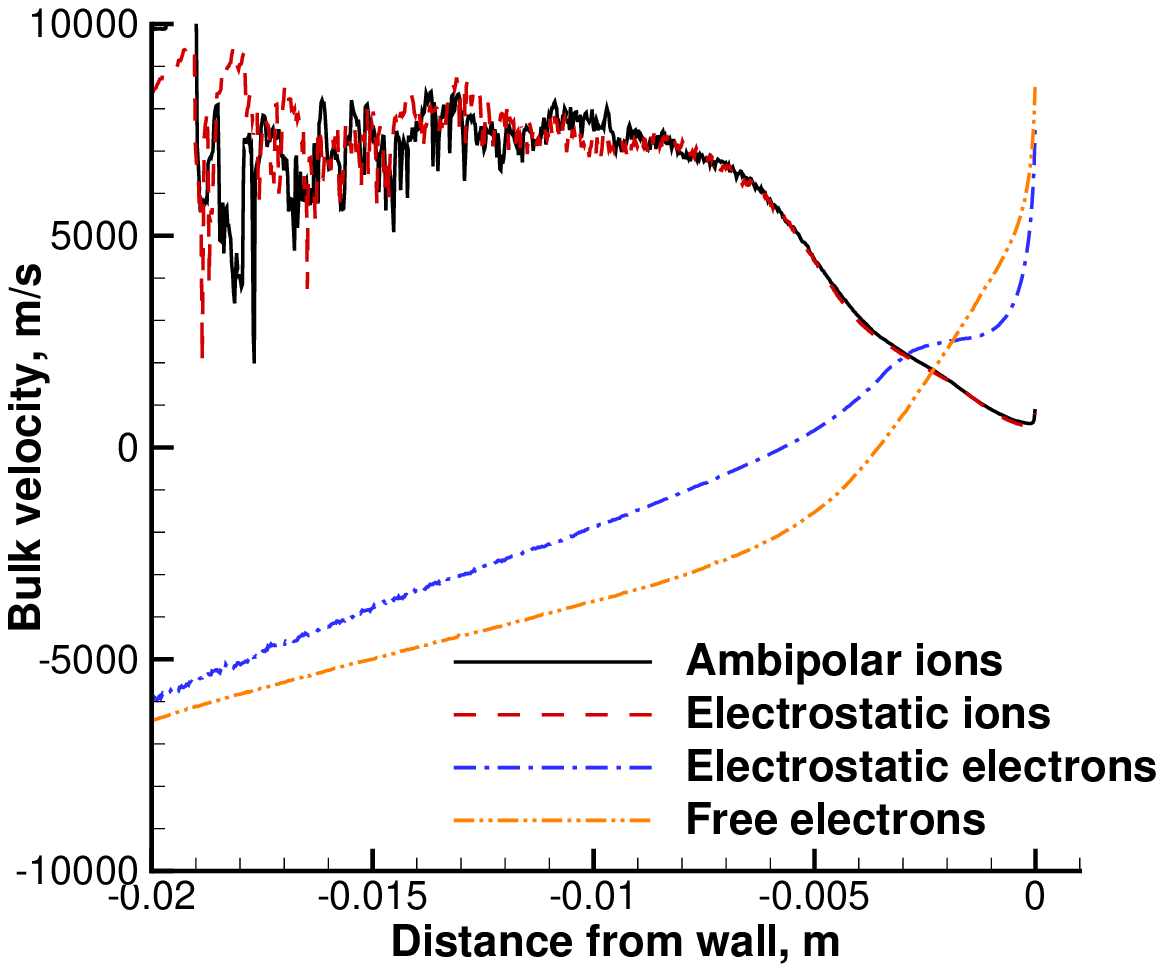}
			\caption{Free diffusion conditions with $F_r = 10^{10}$.}
			\label{fig:60km_Fr=1e10_velocities}
		\end{subfigure}
		\begin{subfigure}{0.49\linewidth}
			\includegraphics[width=\linewidth]{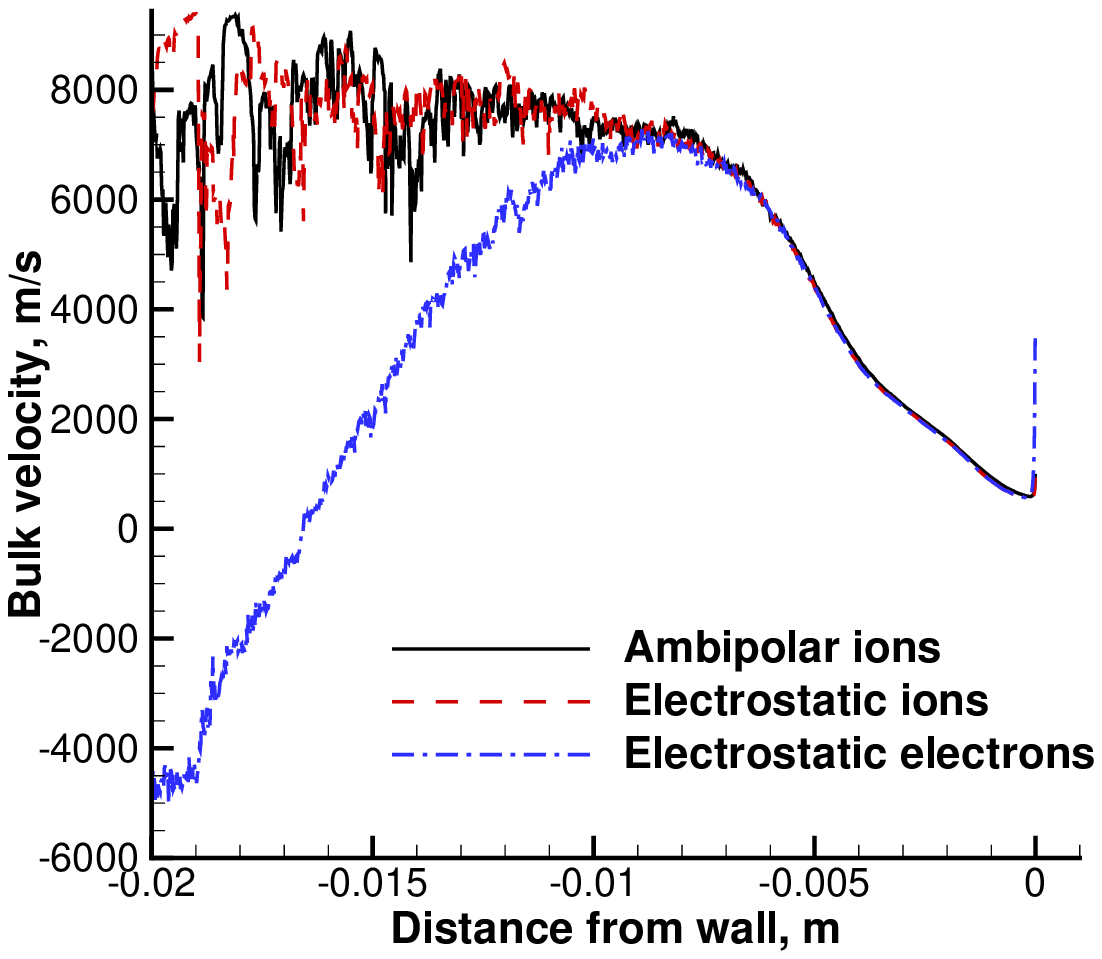}
			\caption{Ambipolar diffusion conditions with $F_r = 10^6$.}
			\label{fig:60km_Fr=1e6_velocities}
		\end{subfigure}
		\caption{Ion and electron bulk velocities under free and ambipolar plasma diffusion conditions for the Mach 35 flow at 60 km.}
	\end{figure*}

	The final point warrants discussion of whether this study's specific implementation of the ambipolar diffusion approximation impacts agreement with electrostatic modeling. An alternative version of the ambipolar diffusion approximation for DSMC advects electrons with the computed ion bulk velocity at each cell\cite{Boyd1997}. This approach does not enforce exact charge neutrality at each cell, unlike the version of the approximation used in this work. Figure \ref{fig:60km_Fr=1e10_velocities} plots the ion and electron bulk velocities for the free diffusion flow case at 60 km with $F_r = 10^{10}$. The ion and electron bulk velocities strongly differ across all models, and thus use of Ref.~\onlinecite{Boyd1997}'s ambipolar diffusion approximation is not expected to alter results. Under ambipolar diffusion conditions at 60 km (Fig.~\ref{fig:60km_Fr=1e6_velocities}), the electrostatic electron and ion bulk velocities align from $x = -0.00725$ to $x = -0.000166$ m. Then, over 20 grid cells, the electron bulk velocity sharply increases to 3,500 m/s at the wall. The large increases in electron convective heat flux across all flow cases are attributed to this rise in electron bulk velocity. Therefore, use of Ref.~\onlinecite{Boyd1997}'s version of the ambipolar diffusion approximation is not expected to alter the conclusions made in this study.
	
	\subsection{Characterization of Plasma Diffusion Regime in Hypersonic Flows}
	\label{sec:summ_character_diff_regime}
	
	Using the `hypersonic diffusion length' (Eq.~(\ref{eq:hypersonic_diff_length})), the criteria for characterizing charged species diffusion regime outlined in Refs.~\onlinecite{Phelps1990} and~\onlinecite{petrusky_evaluation_2026} can successfully be applied towards predicting the plasma diffusion regime in the post-shock region. $\Lambda_h/\lambda_{D,e} \leq 1.0$ is a strong predictor of the free diffusion regime with charge separation on the order of 10\% or more, consistent with the definition from Ref.~\onlinecite{petrusky_evaluation_2026}. $\Lambda_h/\lambda_{D,e} \lesssim 1.0$ is also an effective predictor of the beginning of a plasma sheath, where the $\Lambda_h/\lambda_{D,e}$ profile sharply decreases near the wall. This is imposed by the definition of $\Lambda_h$, but is also consistent with the physics of plasma sheaths which have lengths on the order of several Debye lengths. The exact lower limit of $\Lambda_h/\lambda_{D,e}$ for which a flow enters the ambipolar diffusion regime varies based on simulation setup and gas dynamics modeling. In this study, $\Lambda_h/\lambda_{D,e} \gtrsim 10$ and $\Lambda_h/\lambda_{D,e} \gtrsim 25$ mark the transition into the ambipolar diffusion regime for the electrostatic and ambipolar diffusion approximation models, respectively. For the flow cases of this study, comparing the electron Debye length to the shock standoff distance resulted in similar limits for each plasma diffusion regime; however, use of any constant value prevents observation of a steep drop in $L/\lambda_{D,e}$ when approaching the plasma sheath.

	Using $\Lambda_h/\lambda_{D,e}$ to assess the quasineutrality criterion, the formal definition of the plasma can also be used to characterize hypersonic plasma diffusion regimes. Although the definition only requires $\omega \tau > 1$, this study finds that $\omega_{p,e}\tau_{e,n} < 1$ is a strong predictor of free diffusion and $\omega_{p,e}\tau_{e,n} > 10$ is a strong predictor of ambipolar diffusion with charge separation under 1\%. The plasma parameter $N_D \gg 1$ holds for all flow cases in this study throughout the post-shock region by at least 4 orders of magnitude. Therefore, while $N_D \gg 1$ should always be verified when analyzing a hypersonic plasma, it is not necessarily useful for discriminating between plasma diffusion regimes. 
	
	All flow cases simulated in this study are `strongly' ionized in the post-shock region (i.e., an ionization fraction greater than $10^{-4}$), despite the large variation in plasma densities and diffusion regimes. This highlights that ionization fraction in and of itself is not useful in characterizing charged species diffusion regime unless the total density of the flowfield is known (enabling calculation of quantities such as $\lambda_{D,e}$). Ionization fraction is useful in predicting whether quantities such as stagnation point heat flux may be impacted by electrostatic modeling, since a larger ionization fraction indicates a greater portion of the flowfield will be impacted by plasma effects. 
	
	\subsection{Evaluation of the Ambipolar Diffusion Approximation}
	
	Ultimately, the effectiveness of the ambipolar diffusion approximation in modeling partially ionized hypersonic flows depends on the application and the quantities of interest. If $T_e \ll T_i$, then according to Eq.~\ref{eq:diff_coeff}, the enhanced diffusion rate from self-induced ambipolar electric fields is dominated by the ion diffusion rate. This holds true for the flow cases of this study, where strong agreement is achieved between the ambipolar diffusion approximation's plasma density and the electrostatic ion density. Subsequently, the ambipolar diffusion approximation captures the electrostatic electron density profiles for the ambipolar diffusion flow cases, but not the free and transitional diffusion flow cases. 
		
	The ambipolar diffusion approximation is unable to capture the electrostatic electron temperature profiles across all flow cases. During free diffusion, this is attributed to the independent electron motion leading to differences in electron density distributions that alter collision rates and energy exchange with neutrals. During transitional and ambipolar diffusion, the ambipolar diffusion approximation is unable to capture electron cooling in the vicinity of the wall and electron heating upstream of the shock. This may have important consequences for flows or applications sensitive to electron properties, such as electron-impact ionization-dominated flows or electron transpiration cooling. 
	
	Unlike previous investigations of the ambipolar diffusion approximation in air flows, this work reports decreases in total stagnation point heat flux when electrostatic modeling is used of up to $-7.99\%$. This is attributed to the neutrals losing energy via collisions, leading to their decreased convective heat flux that comprises the majority of heat flux to the wall. This consequence may be unique to the flow conditions and modeling assumptions of this study, but indicates that the tradeoffs of using the ambipolar diffusion approximation in hypersonic flowfield modeling are not universal across all flow conditions and atmospheric compositions. Larger differences between the ambipolar diffusion approximation and electrostatic solutions are reported in the plasma species-specific heat flux quantities, especially the electron convective heat flux. 
	
	\section{Conclusions}
	\label{sec:conclusion}
	
	A series of 1D stagnation streamline simulations were performed across two freestream flow conditions under three different plasma diffusion regimes. Several distinct electrostatic plasma behaviors were identified and discussed across all flow cases, establishing a baseline for understanding plasma behavior for which more complex, chemically reacting partially ionized gas mixtures can be compared to. Although this study and previous literature have demonstrated that the ambipolar diffusion approximation does not accurately capture all flowfield properties, especially the electron temperature and vehicle surface heat flux, the ambipolar diffusion approximation will likely remain the most common approach for simulating a hypersonic plasma due to its computational efficiency. Using criteria such as $\Lambda_h/\lambda_{D,e}$ and $\omega_{p,e}\tau_{e,n}$ with the limits identified in this study, the tradeoffs in using the ambipolar diffusion approximation can be better understood or extrapolated for a given flowfield. Critically, the values of $N_D$ and $\omega_{p,e}\tau_{e,n}$ are similar in magnitude between the ambipolar diffusion approximation and electrostatic solutions within the shock layer; this demonstrates both models are consistent when it comes to predicting whether the partially ionized gas meets the definition of a plasma. As for $\Lambda_h/\lambda_{D,e}$, both models are generally consistent with respect to applying the limits presented in Table \ref{tab:phelps_crit}. Therefore, an ambipolar diffusion approximation solution can be reasonably used to predict the plasma diffusion regime in the post-shock region. An informed decision can then be made on whether high-fidelity electrostatic modeling is required to accurately capture quantities of interest. 
		
		\begin{acknowledgments}
			This material is based upon work supported by the National Science Foundation Graduate Research Fellowship Program under Grant No. DGE 2040434. Any opinions, findings, and conclusions or recommendations expressed in this material are those of the author and do not necessarily reflect the views of the National Science Foundation.
			
			This work utilized the Alpine and Blanca high performance computing resource at the University of Colorado Boulder. Alpine is jointly funded by the University of Colorado Boulder, the University of Colorado Anschutz, and Colorado State University. Blanca is jointly funded by computing users and the University of Colorado Boulder.
		\end{acknowledgments}
		
		\section*{AUTHOR DECLARATIONS}
		
		\subsection*{Conflicts of Interest}
		
		The authors have no conflicts to disclose.
		
		\section*{Data Availability Statement}
		
		The data that support the findings of this study are available from the corresponding author upon reasonable request.

		\appendix
		
		\section{Argon Parameters}
		\label{sec:appendix_argon}
		
		The following appendix provides the details and parameters for the collision and ionization models used throughout this work. 
		
		\begin{table}[h]
			\caption{\label{tab:argon_coll_cross}Momentum exchange collision cross section models used for each reaction pair.}
			\centering
			\begin{tabular}{lc}
				\hline
				\hline
				Collision pair & Model \\
				\hline
				Ar + Ar & Variable Hard Sphere\\
				Ar + Ar$^+$ & Sakabe and Izawa \cite{Sakabe1992}\\
				Ar + e$^-$ & IST-Lisbon Database \cite{lxcat_lisbon,Alves2014} \\
				\hline 
				\hline
			\end{tabular}
		\end{table}
		
		\begin{table}
			\caption{\label{tab:argon_vhs_param}Collision pair parameters used in the variable hard sphere model.}
			\centering
			\begin{tabular}{lcccc}
				\hline
				\hline
				Collision pair & Reference diameter, m& $\omega$ & Reference temperature, K & Source \\
				\hline
				Ar + Ar & 4.17 $\times 10^{-10}$ & 0.81 & 273  & Ref.~\onlinecite{BirdTB1994} \\
				Ar + Ar$^+$ & 4.17 $\times 10^{-10}$ & 0.81 & 273 & Ref.~\onlinecite{BirdTB1994}\\
				Ar + e$^-$ & 1.00 $\times 10^{-10}$ & 0.70 & 273 & Ref.~\onlinecite{Farbar2010}\\
				\hline
				\hline
			\end{tabular}
		\end{table}
		
		\begin{table}
			\caption{\label{tab:argon_ionization_param}Baseline reaction rate coefficients used in the Total Collision Energy chemistry model.}
			\centering
			\begin{tabular}{lcc}
				\hline
				\hline
				Reaction & Rate coefficient, $m^3/s$ & Source \\
				\hline
				Ar + Ar $\rightarrow$ Ar + Ar$^+$ + e$^-$ & 8.9169 $\times 10^{-23} T^{1.1935} exp(-129450/T)$ & Ref.~\onlinecite{drawin_atom-atom_1973}\\
				e$^-$ + Ar $\rightarrow$ Ar$^+$ + 2e$^-$ & 1.23 $\times 10^{-19} T^{1.511} exp(-141480/T)$ & Ref.~\onlinecite{Annaloro2012}\\
				\hline
				\hline
			\end{tabular}
		\end{table}
		
		\bibliography{aipsamp}
		
	\end{document}